\documentclass[12pt]{amsart}
\usepackage{amsmath,amssymb,cite,enumerate,graphicx,hyperref}

\hypersetup{pdftex,colorlinks=true,allcolors=blue}
\usepackage[all]{hypcap}

\usepackage[foot]{amsaddr}
\usepackage[sc]{mathpazo}

\newcommand\ack{\subsection*{Acknowledgment}}

\DeclareMathAlphabet\mathsfbi{T1}{phv}{b}{it}

\numberwithin{equation}{section}

\newcommand\BV{\boldsymbol} 
\newcommand\BM{\mathsfbi} 

\newcommand\dif{\,\mathrm{d}}
\newcommand\deriv[2]{\frac{\dif #1}{\dif #2}}
\newcommand\parderiv[2]{\frac{\partial #1}{\partial #2}}
\newcommand\coll{\mathcal C}
\renewcommand\div{\mathrm{div}}
\newcommand\EE{\mathbb E}
\newcommand\trace{\mathrm{tr}}

\newcommand\eqdef{\stackrel{\text{def}}=}

\newcommand\Pran{\mbox{\textit{Pr}}}

\newcommand\im{\nu}

\newcommand\cL{\mathcal L}
\newcommand\cN{\mathcal N}
\newcommand\cW{\mathcal W}

\newcommand\Vd{W}
\newcommand\Xd{Y}

\newcommand\myatop[2]{\genfrac{}{}{0pt}{}{#1}{#2}}

\begin{document}

\author[Rafail V. Abramov]{Rafail V. Abramov}

\address{Department of Mathematics, Statistics and Computer Science,
University of Illinois at Chicago, 851 S. Morgan st., Chicago, IL 60607}

\email{abramov@uic.edu}

\date{\today}

\title[Diffusive Boltzmann equation and its fluid dynamics]%
{Diffusive Boltzmann equation, its fluid dynamics, Couette flow
and Knudsen layers}

\begin{abstract}
In the current work we construct a multimolecule random process which
leads to the Boltzmann equation in the appropriate limit, and which is
different from the deterministic real gas dynamics process. We
approximate the statistical difference between the two processes via a
suitable diffusion process, which is obtained in the multiscale
homogenization limit. The resulting Boltzmann equation acquires a new
spatially diffusive term, which subsequently manifests in the
corresponding fluid dynamics equations. We test the Navier-Stokes and
Grad closures of the diffusive fluid dynamics equations in the
numerical experiments with the Couette flow for argon and nitrogen,
and compare the results with the corresponding Direct Simulation Monte
Carlo (DSMC) computations. We discover that the full-fledged Knudsen
velocity boundary layers develop with all tested closures when the
viscosity and diffusivity are appropriately scaled in the vicinity of
the walls. Additionally, we find that the component of the heat flux
parallel to the direction of the flow is comparable in magnitude to
its transversal component near the walls, and that the nonequilibrium
Grad closure approximates this parallel heat flux with good accuracy.
\end{abstract}


\maketitle

\section{Introduction}

In the kinetic theory, the processes in gases are described by the
Boltzmann equation \cite{Cer,Cer2,Cer3,CerIllPul}, which models the
evolution of the density~$f(t,\BV x,\BV v)$ of a probability
distribution of a single gas molecule in the space of coordinate~$\BV
x$ and velocity~$\BV v$ at time~$t$, under the assumption that all gas
molecules are independently and identically distributed, and that no
more than two molecules collide at once. The Boltzmann equation is
given by
\begin{equation}
\label{eq:boltzmann}
\parderiv ft+\BV v\cdot\nabla_{\BV x}f=\coll(f),
\end{equation}
where~$\coll(f)$ is the collision term (also called the collision
operator \cite{Lev,Gols}), specified by
\begin{equation}
\label{eq:boltzmann_collision}
\coll(f)=\int B(|(\BV w-\BV v)\cdot\BV n|)\big(f(\BV v')f(\BV w')-f(\BV
v)f(\BV w)\big)\dif\BV n\dif\BV w.
\end{equation}
Above,~$\BV n$ is a unit vector, the integration in $\dif\BV n$ is
over a unit sphere, $B$ is the collision kernel, and~$\BV v'$, $\BV
w'$ are defined by the energy and momentum conservation relations
\begin{equation}
\label{eq:vw}
\BV v'=\BV v+((\BV w-\BV v)\cdot\BV n)\BV n,\qquad \BV
w'=\BV w+((\BV v-\BV w)\cdot\BV n)\BV n.
\end{equation}
Integrating the Boltzmann equation over various powers of the velocity
variable yields the hierarchy of the fluid dynamics equations (also
called the moment equations in kinetic theory), of which the
lowest-order closure is provided by the well known Euler equations
\cite{Bat}, while the next-order closure is given by the Grad
equations \cite{Gra,Gra2}.

The Boltzmann equation and its corresponding hierarchy of the moment
fluid dynamics equations are of the first order in space, which makes
them poorly suitable for boundary value problems.  However, various
formulations of boundary conditions are rather ubiquitous in the
applied problems for realistic flows (for example, the Couette and
Poiseuille flows). For the fluid dynamics equations, the second order
in space is usually achieved via the Chapman-Enskog perturbation
expansion \cite{ChaCow,Gols,Lev}, which, when applied to the Euler
closure, leads to the famous Navier-Stokes equations \cite{Bat}, for
which boundary value problems are usually well posed.

Recently, there appeared a number of works on the extended Boltzmann
equation and the corresponding fluid dynamics
\cite{Brenner,Brenner2,Brenner3,DadRee,Durst,Durst2,Kli,Sher}, where a
spatial diffusion term was introduced in addition to the collision
operator or viscous terms. Some of these works
\cite{Brenner,Brenner2,Brenner3} were based on the idea of introducing
the concept of the ``volume velocity'', which differs from the usual
mass velocity by a small flux term derived from Fick's law. Others
\cite{Durst,Durst2} introduced similar additional terms to model the
self-diffusion of mass. Among those listed above, the works
\cite{DadRee,Kli,Sher} appear to employ the closest conceptual
approach to what we suggest here, namely, in \cite{DadRee,Kli,Sher} an
{\em ad hoc} diffusion term was introduced directly into the Boltzmann
equation~\eqref{eq:boltzmann} via the assumption of an additional
empirical stochasticity of the molecular motion to complement the
already present inter-molecule collisions. However, the main
conceptual drawback of such an approach is its seeming absence of a
fundamental justification. In particular, it was noted in
\cite{DadRee} that there did not seem to be a thermodynamically valid
reason to introduce an additional diffusion term into a continuum gas
model.

In the current work, we identify the difference between the actual
dynamics of a real gas, and the dynamics which are described by the
Boltzmann equation in~\eqref{eq:boltzmann}. More specifically, we
construct the precise multimolecular dynamical system which leads
directly to the Boltzmann equation in the appropriate limit, and this
system turns out to be fundamentally different from the realistic gas
dynamics. While the real gas dynamics is fully deterministic with the
molecular trajectories prescribed exactly via the initial conditions,
the multimolecular system we introduce is a random jump process (more
precisely, a L\'evy-type Feller process~\cite{App}), whose
stochasticity is inherent. Then, we approximate the difference between
the two dynamics via a multscale formalism in the homogenization time
limit~\cite{KelMel,Kif,PavStu,Van,Vol}, which equips the Boltzmann
equation and the corresponding fluid dynamics equations with an
additional spatially diffusive term.

The manuscript is organized as follows. In Section~\ref{sec:real_gas}
we consider the multimolecular motion of a realistic gas where the
intermolecular interactions are governed by a repelling potential, and
arrive at the conclusion that the deterministic multimolecular
Liouville equation cannot be directly simplified into the Boltzmann
equation without an appropriate stochastic modification. In
Section~\ref{sec:random_jumps} we construct the aforementioned
multimolecular random jump process, whose Kolmogorov equation
naturally reduces to the Boltzmann equation in the appropriate
limit. In Section~\ref{sec:multiscale} we approximate the difference
between the two multimolecular systems in the multiscale
homogenization time limit, which leads to an It\^o diffusion
process. In Section~\ref{sec:diffusive_boltzmann} we augment the
multimolecular random jump process with the corrective It\^o diffusion
process, such that the resulting Boltzmann equation acquires a spatial
diffusion term, which corresponds to the diffusion of mass. In
Section~\ref{sec:fluid_dynamics} we derive the diffusive Euler,
Navier-Stokes, Grad \cite{Gra,Gra2} and regularized Grad
\cite{Stru,StruTor,TorStru} equations, which also inherit the spatial
diffusion terms. In Section~\ref{sec:numerics} we carry out the
numerical simulations for argon and nitrogen in the simple Couette
flow setting, and compare them against the DSMC
computations~\cite{Bird,ScaRooWhiDarRee}. We discover that if the
viscosity and diffusivity are appropriately scaled in the vicinity of
the walls~\cite{Abr15}, then all studied moment closures develop the
full-fledged Knudsen velocity boundary layers in agreement with the
DSMC computations. We also find that the diffusive Navier-Stokes
equations produce a better temperature prediction than the
conventional Navier-Stokes equations.  We observe that the DSMC
computations produce a substantial component of the heat flux parallel
to the direction of the flow, which the Fourier law approximation of
the Navier-Stokes equations fails to capture. On the other hand, both
the diffusive and regularized diffusive Grad closures approximate the
parallel heat flux, the latter with particularly good accuracy. The
summary is given in Section~\ref{sec:summary}.

\section{The microscopic dynamics of a realistic gas}
\label{sec:real_gas}

Here we consider a system of~$K$ identical gas molecules, which move
in an $N$-dimensional Euclidean space. For a realistic gas, the
components $i=1, 2, 3$ of the space are translational, while
$i=4,\ldots,N$ are rotational, and thus naturally periodic. Also, the
way real gas molecules interact depends differently on the
translational and rotational coordinates; clearly, the translational
coordinates of a pair of molecules have to be similar for them to
interact, while the alignment of their rotational coordinates weakly
affect the occurrence or non-occurrence of a collision (but not its
outcome, obviously).

For convenience, below we will treat all coordinates equivalently, as
both translational, as far as the molecular interaction is concerned,
and rotational, as we assume that the full coordinate space is
Euclidean and periodic in each coordinate. In this sense, any gas we
consider below is, in a way, ``monatomic'', except that its phase
space can be more than three-dimensional. While such an approach does
not rigorously address the issues of collision between polyatomic gas
molecules, below we will show that it is not needed for our purposes
-- we will fully decouple the rotational coordinates from the
macroscopic equations of the gas dynamics. Besides, the collision of
polyatomic molecules is a separate and rather complex subject in
itself, and often phenomenological models (such as the
Borgnakke-Larsen collision model \cite{BorLar}) are used to describe
such collisions in practice.

In order to define a multimolecular dynamical system in a concise
manner, we concatenate all coordinates~$\BV x_i$ and velocities~$\BV
v_i$ of the individual molecules into the two vectors~$\BV X$ and~$\BV
V$ as follows:
\[
\BV X=\left(\BV x_1,\BV x_2,\ldots,\BV x_K\right),\qquad \BV
V=\left(\BV v_1,\BV v_2,\ldots,\BV v_K\right).
\]
Apparently,~$\BV X$ and~$\BV V$ have the dimension~$KN$, since
each~$\BV x_i$ or~$\BV v_i$ is $N$-dimensional.

We assume that the molecules interact with each other via a potential
function $H(\BV X)$, so that the equations of motion for the realistic
gas molecules are given by
\begin{equation}
\label{eq:dyn_sys_real}
\deriv{\BV X}t=\BV V,\qquad\deriv{\BV V}t=-\nabla_{\BV X}H(\BV X).
\end{equation}
In what follows, we will assume that the potential $H$ has the form of
the sum of pairwise interactions between all molecules in the system:
\begin{equation}
\label{eq:H}
H(\BV X)=\sum_{\myatop{i=1\ldots K-1}{j=i+1\ldots K}}\phi(\|\BV
x_i-\BV x_j\|),
\end{equation}
where $\phi(\|\BV x\|)$ is the potential of an individual molecule
(for example, the Lennard-Jones potential \cite{Len}). It is easy to
see that such a choice of $H$ ensures the conservation of the total
momentum $\BV m$ and energy $E$ of the system,
\begin{equation}
\label{eq:mom_en}
\BV m=\sum_{i=1\ldots K}\BV v_i,\qquad E=\frac 12\|\BV V\|^2+H(\BV X).
\end{equation}
Let $F(t,\BV X,\BV V)$ be the density distribution function of $\BV X$
and $\BV V$. Then, the evolution equation for $F$ is given by the
corresponding Liouville equation (which is also the special case of
the forward Kolmogorov equation~\cite{GikSko} for a deterministic
system) for the real gas:
\begin{equation}
\label{eq:liouville}
\parderiv Ft+\BV V\cdot\nabla_{\BV X}F=\nabla_{\BV X}H\cdot\nabla_{\BV
  V}F.
\end{equation}
The Liouville equation in~\eqref{eq:liouville} admits a special class
of solutions in the form of the product of identical probability
densities $f$ for each molecule,
\begin{equation}
\label{eq:F_prod}
F(t,\BV X,\BV V)=\prod_{i=1\ldots K}f(t,\BV x_i,\BV v_i).
\end{equation}
Indeed, substituting~\eqref{eq:F_prod} into~\eqref{eq:liouville}, we
obtain, for the different terms of~\eqref{eq:liouville},
\begin{subequations}
\label{eq:F_prod_2}
\begin{equation}
\parderiv Ft=\sum_{i=1}^K\left(\prod_{j\neq i}f(t,\BV x_j,\BV v_j)
\right)\parderiv{}tf(t,\BV x_i,\BV v_i),
\end{equation}
\begin{equation}
\BV V\cdot\nabla_{\BV X}F=\sum_{i=1}^K\left(\prod_{j\neq i}f(t,\BV
x_j,\BV v_j)\right)\BV v_i\cdot\nabla_{\BV x}f(t,\BV x_i,\BV v_i),
\end{equation}
\begin{equation}
\nabla H(\BV X)\cdot\nabla_{\BV V}F=\sum_{i=1}^K\sum_{j\neq i}\bigg(
\prod_{\myatop{k\neq i}{k\neq j}}f(t,\BV x_k,\BV v_k)\bigg)f(t,\BV
x_j,\BV v_j)\nabla_{\BV x_i}\phi(\|\BV x_i-\BV x_j\|)\cdot\nabla_{\BV
  v} f(t,\BV x_i,\BV v_i).
\end{equation}
\end{subequations}
Substituting~\eqref{eq:F_prod_2} into~\eqref{eq:liouville} and
integrating over $\BV x_i$, $\BV v_i$, $i>1$, under the assumption
that all $\BV x_i$- and $\BV v_i$-derivatives integrate out to zero we
arrive at an equation for the distribution $f(t,\BV x,\BV v)$ of a
single molecule in the closed form, known as the Vlasov equation:
\begin{equation}
\label{eq:vlasov}
\parderiv ft+\BV v\cdot\nabla_{\BV x}f=\nabla_{\BV x}h\cdot
\nabla_{\BV v}f,
\end{equation}
where $h(\BV x)$ is the combined potential of the remaining $(K-1)$
identical molecules:
\begin{equation}
\label{eq:h}
h(\BV x)=(K-1)\int\phi(\|\BV x-\BV y\|)f(\BV y,\BV w)\dif\BV y\dif\BV
w.
\end{equation}
Thus, clearly, if $f(t,\BV x,\BV v)$ is a solution
of~\eqref{eq:vlasov}, then the product~\eqref{eq:F_prod} is a solution
of~\eqref{eq:liouville}.

Observe that in the limit as $K$ becomes large, $h(\BV x)$ grows in an
unbounded fashion. This happens because the finite volume of the
periodic $N$-dimensional Euclidean space for the molecular coordinates
$\BV x$ is fixed while more molecules are added into it, thus
decreasing the mean distance between the molecules. To counter this
effect, one can appropriately rescale the range of the interaction
potential $\phi$ to fix the ratio of the mean intermolecular distance
to the interaction range, which is known as the Boltzmann-Grad limit
\cite{Gra,Gra2}.

Observe that the right-hand side of the Vlasov equation
in~\eqref{eq:vlasov} contains the derivative of $f$ with respect to
the velocity $\BV v$, whereas the right-hand side of the Boltzmann
equation in~\eqref{eq:boltzmann} contains, on the contrary, the
integrals of $f$ with respect to the velocity (which are a
manifestation of randomness in the dynamical process). Below we
introduce the exact multimolecular random dynamical process, which is
different from~\eqref{eq:dyn_sys_real} and leads to the Boltzmann
equation in~\eqref{eq:boltzmann}, rather than the Vlasov equation
in~\eqref{eq:vlasov}.

\section{A stochastic modification of the Liouville equation which
leads to the Boltzmann equation}
\label{sec:random_jumps}

In what follows, we replace the deterministic intermolecular
interaction term in the Liouville equation~\eqref{eq:liouville} with
the stochastic interaction term of our choice. We then show that the
same procedure which leads from~\eqref{eq:liouville}
to~\eqref{eq:vlasov}, results in the Boltzmann equation for the newly
introduced random dynamics.

Observe that the interaction in the right-hand side of Liouville
equation in~\eqref{eq:liouville} can be written in the form
\begin{multline}
\label{eq:liouville_rhs}
\nabla_{\BV X}H\cdot\nabla_{\BV V}F(t,\BV X,\BV V)=\left.\parderiv{}s
F(t,\BV X,\BV V+s\nabla_{\BV X}H)\right|_{s=0}=\\=\lim_{s\to 0}\frac
1s \left[F(t,\BV X,\BV V+s\nabla_{\BV X}H)-F(t,\BV X,\BV V)\right]=
\\=\lim_{s\to 0}\frac 1s \left[\int G_s(\BV X;\BV V'|\BV V)F(t,\BV
  X,\BV V') \dif\BV V'-F(t,\BV X,\BV V)\right]=\\=\lim_{s\to 0}\frac
1s \int G_s(\BV X;\BV V'|\BV V)\left[F(t,\BV X,\BV V')-F(t,\BV X,\BV
  V)\right] \dif\BV V'.
\end{multline}
Above, the formally introduced conditional probability density
$G_s(\BV X;\BV V'|\BV V)$ is completely deterministic and merely
specifies that $\BV V'=s\nabla_{\BV X}H+\BV V$:
\begin{equation}
\label{eq:Gs}
G_s(\BV X;\BV V'|\BV V)=\delta(s\nabla_{\BV X}H+\BV V-\BV V'),
\end{equation}
where ``$\delta$'' is Dirac's delta-function. In order to arrive at
the Boltzmann equation, we will replace the deterministic conditional
probability density $G_s$ in~\eqref{eq:Gs} with a stochastic
alternative $G_s^B$ (where ``$B$'' stands for ``Boltzmann'') of our
choosing. However, observe that we do not need $G_s^B$ directly;
instead, specifying its $s$-derivative at $s=0$ will suffice.
Following the structure of the potential $H$ in~\eqref{eq:H}, we
choose the new conditional density to be the sum of all pairwise
conditional molecular interactions, depending on the molecule
coordinates:
\begin{multline}
\label{eq:GB}
\left.\parderiv{}sG_s^B(\BV X;\BV V'|\BV V)\right|_{s=0}\eqdef G^B(\BV
X;\BV V'|\BV V)=\\=\sum_{ \myatop{i=1\ldots K-1}{j=i+1 \ldots K}}
\bigg(g(\BV v_i',\BV v_j'|\BV v_i,\BV v_j)\chi_d\left(\|\BV x_i-\BV
x_j\|\right)\prod_{\myatop{k\neq i}{k\neq j}}\delta(\BV v_k'-\BV v_k)
\bigg).
\end{multline}
Above, $\chi_d$ is the indicator function of a ball of diameter $d$,
which is the size of a gas molecule (so that $\chi_d(\|\BV x_i-\BV
x_j\|)=1$ whenever $\|\BV x_i-\BV x_j\|\leq d$, and zero
otherwise). Observe that $G^B$ is not normalized to one; instead, its
norm (also called ``activity'') controls the overall time rate of
change of $F$.

With help of the Boltzmann collision kernel $B$
from~\eqref{eq:boltzmann_collision}, we define the two-molecule
conditional density $g$ as
\begin{multline}
\label{eq:g}
g(\BV v',\BV w'|\BV v,\BV w)=\frac 1{V_N(d)K} B\big(\|\BV v'-\BV
w'+\BV w- \BV v\|/2\big)\times\\\times\delta(\BV v'+\BV w'-\BV v-\BV
w)\delta \left(\|\BV v'\|^2+\|\BV w'\|^2-\|\BV v\|^2-\|\BV
w\|^2\right),
\end{multline}
where $V_N(d)$ is the volume of an $N$-dimensional ball of diameter
$d$. The coefficient $(V_N(d)K)^{-1}$ in front of $B$ above
in~\eqref{eq:g} ensures the correct scaling of the activity of $g$ in
the Boltzmann-Grad limit \cite{Gra,Gra2}, where the ratio of the mean
distance between the molecules to the molecular diameter has to be
fixed in the limit as the number of molecules $K\to\infty$. We imply
that the molecular diameter $d$ is adjusted accordingly with
increasing $K$, so that the product $V_N(d)K$ approaches a finite
value as $K\to\infty$.

The modified Liouville equation in~\eqref{eq:liouville} thus becomes
the forward Kolmogorov equation in the form
\begin{equation}
\label{eq:gen_kolm}
\parderiv Ft+\BV V\cdot\nabla_{\BV X}F=\int G^B(\BV X;\BV V'|\BV V)
\left[F(t,\BV X,\BV V')-F(t,\BV X,\BV V)\right] \dif\BV V'.
\end{equation}
Just as with the Liouville equation~\eqref{eq:liouville}, one can
choose a solution $F$ of~\eqref{eq:gen_kolm} in the factored
form~\eqref{eq:F_prod}. However, in this case the single-molecule
density $f(t,\BV x,\BV v)$ satisfies the Boltzmann
equation~\eqref{eq:boltzmann} in the Boltzmann-Grad limit. Indeed, let
us substitute~\eqref{eq:F_prod} into~\eqref{eq:gen_kolm}, obtaining
the same relations as in~\eqref{eq:F_prod_2}, with the exception of
the right-hand side of~\eqref{eq:gen_kolm}, which is given by
\begin{multline*}
\int G^B(\BV X;\BV V'|\BV V) \left[F(t,\BV X,\BV V')-F(t,\BV X,\BV
  V)\right] \dif\BV V'=\\=\sum_{ \myatop{i=1\ldots K-1}{j=i+1 \ldots
    K}}\bigg(\prod_{\myatop{k\neq i}{k\neq j}} f(t,\BV x_k,\BV v_k)
\bigg)\int g(\BV v_i',\BV v_j'|\BV v_i,\BV v_j)\chi_d\left(\|\BV x_i
-\BV x_j\|\right)\times\\\times\left(f(t,\BV x_i,\BV v_i')f(t,\BV x_j,
\BV v_j')-f(t,\BV x_i,\BV v_i)f(t,\BV x_j,\BV v_j)\right)\dif\BV v_i'
\dif\BV v_j'.
\end{multline*}
As for the Vlasov equation in~\eqref{eq:vlasov}, integrating over $\BV
x_i$, $\BV v_i$, $i>1$, we arrive at the equation for $f$ in the
closed form:
\begin{multline*}
\parderiv{}tf(t,\BV x,\BV v)+\BV v\cdot\nabla_{\BV x}f(t,\BV x,\BV v)
=(K-1)\int g(\BV v',\BV w'|\BV v,\BV w)\chi_d\left(\|\BV x-\BV y\|
\right)\times\\\times\left(f(t,\BV x,\BV v')f(t,\BV y,\BV w') -f(t,\BV
x,\BV v)f(t,\BV y,\BV w)\right)\dif\BV v'\dif\BV w'\dif\BV y\dif\BV w.
\end{multline*}
Now we assume that the molecular diameter $d$ is small enough (which
means that the number of molecules $K$ is large enough), so that the
integration over $\chi_d(\|\BV x-\BV y\|)\dif\BV y$ can be replaced
with setting $\BV y=\BV x$ and multiplying by $V_N(d)$. This further
yields
\begin{multline*}
\parderiv{}tf(t,\BV x,\BV v)+\BV v\cdot\nabla_{\BV x}f(t,\BV x,\BV v)
=V_N(d)(K-1)\int g(\BV v',\BV w'|\BV v,\BV w)\times\\\times\left(f(t,
\BV x,\BV v')f(t,\BV x,\BV w') -f(t,\BV x,\BV v)f(t,\BV x,\BV w)
\right)\dif\BV v'\dif\BV w'\dif\BV w.
\end{multline*}
Finally, substituting $g$ from~\eqref{eq:g} and observing that
$(K-1)/K\approx 1$, we obtain the Boltzmann equation
in~\eqref{eq:boltzmann}.

It remains to determine the underlying dynamics equation which
produces the forward Kolmogorov equation in~\eqref{eq:gen_kolm}. One
can verify that it is the Langevin equation \cite{Lan} in a
generalized sense, given by
\begin{equation}
\label{eq:dyn_sys_alt}
\deriv{\BV X}t=\BV V,\qquad\dif\BV V=\dif\BV\cN(t).
\end{equation}
Above, $\BV\cN(t)$ is a random jump process whose generator is given,
for a suitable function $\psi(\BV V)$, by
\[
\parderiv{}t\EE\psi=\int\big(\psi(\BV V')-\psi(\BV V)\big)G^B(\BV X;
\BV V'|\BV V)\dif\BV V',
\]
as verified via integration by parts. Recalling Courr\`ege's theorem
\cite{Cou}, we conclude that the full process $(\BV X(t),\BV V(t))$ is
a L\'evy-type Feller process \cite{App}. The random dynamics
in~\eqref{eq:dyn_sys_alt} preserve the momentum and kinetic energy,
given by
\begin{equation}
\label{eq:mom_en_B}
\BV m=\sum_{i=1}^K\BV v_i,\qquad E=\frac 12\|\BV V\|^2.
\end{equation}

\subsection*{Microcanonical and canonical Gibbs ensembles}

From~\eqref{eq:mom_en_B}, it follows that the process $\BV V(t)$ lives
on the $((K-1)N-1)$-dimensional sphere of constant momentum $\BV m$
and kinetic energy $E$, given by~\eqref{eq:mom_en_B}, and, in
particular, has a stationary distribution which is uniform on this
sphere. Such a solution is known as the microcanonical Gibbs ensemble
\cite{Gib}. On the other hand, the product representation
in~\eqref{eq:F_prod} cannot be chosen to be such a microcanonical
distribution because it cannot be supported on a closed
$((K-1)N-1)$-dimensional sphere and nowhere else due to its
structure. Instead, we know from the theory for the Boltzmann equation
\cite{Cer,CerIllPul,Gols}, that its stationary solution is a Gaussian
distribution with variance $\theta$ (also known as the temperature),
and thus the corresponding product~\eqref{eq:F_prod} of the stationary
Boltzmann distributions is given by
\begin{equation}
F=C\exp(-E/\theta),
\end{equation}
where $C$ is an appropriate normalization coefficient. This type of
solution of~\eqref{eq:gen_kolm} is known as the canonical Gibbs
ensemble. Note that the microcanonical and canonical Gibbs states are
completely different solutions of~\eqref{eq:gen_kolm}.

However, observe that, for large $K$, the two-molecule marginal of a
uniform distribution on the $((K-1)N-1)$-dimensional constant
momentum/energy sphere becomes the product of two independent and
identical Gaussian distributions. Indeed, let us fix the velocities of
two first molecules $\BV v_1$ and $\BV v_2$. Then, the surface area of
the sphere occupied by the rest of the molecules is proportional to
\begin{multline*}
S\sim\left(KN\theta-(\|\BV v_1\|^2+\|\BV v_2\|^2)
\right)^{((K-3)N-1)/2} \sim\left(1-\frac{\|\BV v_1\|^2+\|\BV v_2\|^2
}{KN\theta} \right)^{((K-3)N-1)/2}=\\=\left(1-\frac 2{KN}\frac{\|\BV
  v_1\|^2+\|\BV v_2\|^2}{ 2\theta}\right)^{KN/2-(3N+1)/2}\sim\exp
\left(-\frac{\|\BV v_1\|^2}{2\theta}\right)\exp\left(-\frac{\|\BV
  v_2\|^2}{2\theta}\right),
\end{multline*}
that is, we arrive at the product of two independent Gaussian
distributions for $\BV v_1$ and $\BV v_2$ with the identical
temperature $\theta$ as $K\to\infty$. Thus, the marginal distributions
of the microcanonical and canonical Gibbs ensembles at statistical
equilibrium are equivalent for large $K$.

Now, let $F$ be a microcanonical ensemble solution
of~\eqref{eq:gen_kolm} where the molecules are statistically
indistinguishable (that is, $F$ is invariant under arbitrary
renumbering of the molecules), and let $f^{(1)}$ and $f^{(2)}$ denote
its single- and two-molecule marginals, respectively. Then, as before,
integrating~\eqref{eq:gen_kolm} over $\BV x_i$, $\BV v_i$, $i>1$, we
arrive at
\begin{multline*}
\parderiv{}tf^{(1)}(t,\BV x,\BV v)+\BV v\cdot\nabla_{\BV x}f^{(1)}
(t,\BV x,\BV v) =(K-1)\int g(\BV v',\BV w'|\BV v,\BV w)\chi_d
\left(\|\BV x-\BV y\| \right)\times\\\times\left(f^{(2)}(t,\BV x,\BV
y,\BV v',\BV w')-f^{(2)}(t,\BV x,\BV y,\BV v,\BV w)\right)\dif\BV
v'\dif\BV w'\dif\BV y\dif\BV w,
\end{multline*}
and, assuming that $f^{(2)}$ separates into the product of two
single-molecule marginals $f^{(1)}$ as above with good accuracy, we
again arrive at the Boltzmann equation~\eqref{eq:boltzmann} in the
Boltzmann-Grad limit.

Thus, in what follows, we will assume that the solution of the
Boltzmann equation in~\eqref{eq:boltzmann} is a valid approximation of
the single-molecule marginal of a microcanonical solution
of~\eqref{eq:gen_kolm} in the Boltzmann-Grad limit, and the main
difference between the statistical behavior of the realistic gas
system in~\eqref{eq:dyn_sys_real} and the solution of the Boltzmann
equation in~\eqref{eq:boltzmann} comes from the statistical difference
in the dynamics of~\eqref{eq:dyn_sys_real} and~\eqref{eq:dyn_sys_alt}
(or, equivalently, from the difference in the corresponding
microcanonical solutions of the Liouville equation
in~\eqref{eq:liouville} and the Kolmogorov equation
in~\eqref{eq:gen_kolm}).

\section{The long-term behavior of a gas}
\label{sec:multiscale}

Observe that the random jump process in~\eqref{eq:dyn_sys_alt} is a
completely different dynamical system from the realistic gas process
in~\eqref{eq:dyn_sys_real}. First, the collision is not guaranteed to
happen even if the two molecules are within the interaction range,
since the random jump event in $\BV\cN(t)$ may not necessarily arrive
during that time (which implies that~\eqref{eq:dyn_sys_alt} ``misses''
some collisions which occur in~\eqref{eq:dyn_sys_real}). Second, when
the collision occurs, the velocity vectors of deflected molecules are
determined at random (as opposed to the deterministic deflections
in~\eqref{eq:dyn_sys_real}), albeit under the momentum and energy
conservation constraints. In what follows, we suggest a statistical
correction to~\eqref{eq:dyn_sys_alt} (and,
therefore~\eqref{eq:gen_kolm}) to better match the dynamics
in~\eqref{eq:dyn_sys_real} in the long-term limit. This correction is
based on the multiscale analysis of the difference dynamics between
the two systems, and manifests in the form of an additional spatial
diffusion term in the Boltzmann equation~\eqref{eq:boltzmann}.

Let $\BV\Xd(t)$ and $\BV\Vd(t)$ denote, respectively,
\begin{equation}
\label{eq:diff}
\BV\Xd(t)=\BV X_{\eqref{eq:dyn_sys_alt}}(t)-\BV
X_{\eqref{eq:dyn_sys_real}}(t),\qquad\BV\Vd(t)=\BV
V_{\eqref{eq:dyn_sys_alt}}(t)-\BV V_{\eqref{eq:dyn_sys_real}}(t),
\end{equation}
and satisfy, respectively,
\begin{equation}
\label{eq:dyn_sys_diff}
\deriv{\BV\Xd}t=\BV\Vd,\qquad\dif\BV\Vd=\nabla H(\BV X)\dif
t+\dif\BV\cN(t),
\end{equation}
where~\eqref{eq:dyn_sys_diff} is coupled to~\eqref{eq:dyn_sys_real}.
We will assume that the momentum and energy of~\eqref{eq:dyn_sys_real}
and~\eqref{eq:dyn_sys_alt} are identical, which, in particular,
implies that the momentum of $\BV\Vd$ is zero.

Our goal here is to obtain an approximate equation for the process
$\BV\Xd(t)$ alone in a suitable ``closed'' form, that is, the one that
does not involve the processes $\BV X(t)$, $\BV V(t)$, and $\BV
\Vd(t)$. For that, we split $\BV\Xd(t)$ into two processes -- the
``macroscale'' process, and the ``microscale'' process. Observe that,
in general, we can distinguish between two spatial scales in the
evolution of a gas system -- the microscale, on which the molecules
move and interact with each other, and the macroscale, on which the
statistical properties of a gas, such as the temperature, vary. For
example, at standard conditions (sea level pressure, room
temperature), the microscale of evolution of a gas is about 60-70
nanometers (the length of the mean free path between molecular
collisions), while significant variations of the temperature are
typically observed on a much larger scale.

Next, following \cite{PavStu}, we introduce the additional macroscale
process $\BV Z(t)=-\varepsilon\BV\Xd(t)$ (the importance of the
opposite sign will manifest later), which is coupled to the difference
process in~\eqref{eq:dyn_sys_diff} (which, in turn, is coupled to the
realistic gas system in~\eqref{eq:dyn_sys_real}) without a feedback
coupling. As $\BV Z(t)$ obviously varies slower than $\BV \Xd(t)$ due
to scaling, we also rescale the time variable $t$ as $\varepsilon^2t$
to obtain $\BV Z(t)$ in the homogenization limit
\cite{PavStu,Van}. The resulting system is given by
\begin{subequations}
\label{eq:dyn_sys_rescaled}
\begin{equation}
\deriv{\BV Z}t=-\frac 1\varepsilon\BV\Vd,\qquad\deriv{\BV\Xd}t
=\frac 1{\varepsilon^2}\BV\Vd,\qquad\dif\BV\Vd=\frac 1{
  \varepsilon^2}\nabla H(\BV X)\dif t+\dif\BV\cN(t/\varepsilon^2),
\end{equation}
\begin{equation}
\deriv{\BV X}t=\frac 1{\varepsilon^2}\BV V,\qquad\deriv{\BV V}t=-\frac
1{\varepsilon^2}\nabla H(\BV X).
\end{equation}
\end{subequations}
Observe that the system above is the one in
\eqref{eq:dyn_sys_real}+\eqref{eq:dyn_sys_diff}, with the additional
macroscale process $\BV Z(t)$, and with the time variable rescaled by
$\varepsilon^2$. Even though $\BV Z(t)$ is not coupled back to the
fast variables $\BV X(t)$, $\BV V(t)$, $\BV\Xd(t)$ or $\BV\Vd(t)$
above in~\eqref{eq:dyn_sys_rescaled}, we assume that the statistical
properties of the fast variables depend on $\BV Z(t)$ in the
homogenization limit.

Before we proceed with the multiscale formalism, we make the following
assumptions about the joint process
\eqref{eq:dyn_sys_real}+\eqref{eq:dyn_sys_diff}: first we will assume
that it is ergodic and strongly mixing with rapid decay of time
autocorrelation functions, and, second, that all the components of
$\BV X(t)$, $\BV V(t)$, $\BV\Xd(t)$ and $\BV\Vd(t)$ are statistically
identical at equilibrium. The latter assumption, combined with the
zero $\BV\Vd$-momentum assumption, implies that the statistical
equilibrium average of $\BV\Vd(t)$ is zero (so-called centering
condition \cite{PavStu}).

We are now going to use the multiscale formalism
\cite{PavStu,Vol,Van,Kif,KelMel} to obtain the approximate closed
dynamics in the long-term limit for $\BV Z(t)$. The corresponding
extended Kolmogorov equation for~\eqref{eq:dyn_sys_rescaled} is given
by
\begin{equation}
\label{eq:kolm_rescaled}
\parderiv {F^{ext}}t-\frac 1\varepsilon\BV\Vd\cdot \nabla_{\BV Z}
F^{ext}=\frac 1{\varepsilon^2}\cL^*F^{ext},
\end{equation}
where $\cL$ is the standalone generator of the difference process
in~\eqref{eq:dyn_sys_diff}, coupled to~\eqref{eq:dyn_sys_real}:
\begin{multline}
\cL^* F^{ext}=\int G^B(\BV X+\BV\Xd;\BV\Vd'|\BV\Vd)(F^{ext}(\BV
\Vd')-F^{ext}(\BV\Vd))\dif\BV\Vd'-\\-\BV\Vd\cdot\nabla_{\BV
 \Xd}F^{ext}-\nabla H(\BV X)\cdot\nabla_{\BV\Vd}F^{ext}+\nabla H(\BV
X)\cdot\nabla_{\BV V}F^{ext}-\BV V\cdot\nabla_{\BV X}F^{ext}.
\end{multline}
To obtain a solution for~\eqref{eq:kolm_rescaled}, we expand $F^{ext}$
as
\[
F^{ext}=F^{ext}_0+\varepsilon F^{ext}_1+\varepsilon^2 F^{ext}_2+\ldots,
\]
where $F^{ext}_0$ has the same normalization as $F^{ext}$, while
$F^{ext}_i$, $i>0$ are normalized to zero. Substituting the expansion
above into~\eqref{eq:kolm_rescaled}, we obtain, in the consecutive
orders of $\varepsilon$,
\begin{subequations}
\begin{equation}
\label{eq:e0}
\cL^*F^{ext}_0=0,
\end{equation}
\begin{equation}
\label{eq:e1}
\cL^*F^{ext}_1=-\BV\Vd\cdot\nabla_{\BV Z} F^{ext}_0,
\end{equation}
\begin{equation}
\label{eq:e2}
\parderiv {F^{ext}_0}t-\BV\Vd\cdot\nabla_{\BV Z}F^{ext}_1
=\cL^*F^{ext}_2.
\end{equation}
\end{subequations}
From~\eqref{eq:e0} it follows that $F^{ext}_0$, suitably normalized,
is an invariant measure for~\eqref{eq:dyn_sys_diff}, coupled
to~\eqref{eq:dyn_sys_real}. In particular, let $\bar F$ denote the
marginal
\begin{equation}
\bar F=\int F^{ext}\dif\BV X\dif\BV V\dif\BV\Xd\dif\BV\Vd,
\end{equation}
then, for an arbitrary $F^{ext}$-measurable function $\psi$,
\begin{equation}
\int\psi F^{ext}_0\dif\BV X\dif\BV V\dif\BV\Xd\dif\BV\Vd=\bar
F_0\int\psi\dif\im,
\end{equation}
where $\im$ is the ergodic invariant measure for the joint system
\eqref{eq:dyn_sys_real}+\eqref{eq:dyn_sys_diff}.

Now, we integrate~\eqref{eq:e2} with respect to $\dif\BV X\dif\BV
V\dif\BV\Xd\dif\BV\Vd$ and find
\begin{equation}
\label{eq:Fext1}
\parderiv{\bar F_0}t=\div_{\BV Z}\int\BV\Vd F^{ext}_1\dif\BV X\dif\BV
V\dif\BV\Xd\dif\BV\Vd,
\end{equation}
as the integral of the term in the right-hand side of~\eqref{eq:e2} is
zero. To evaluate the integral in~\eqref{eq:Fext1} above, let us
denote
\begin{equation}
\Phi(t,\BV V)=\int_0^t\EE\BV\Vd(s)\dif s,\qquad\BV\Vd(0)=\BV W,
\end{equation}
where the expectation $\EE\BV\Vd(t)$ is taken over all realizations of
the random jump process $\BV\cN(t)$. Then, it can be shown (see, for
example, Lemma 3.2.2 in \cite{App}) that
\[
\cL\Phi(t,\BV W)=\EE\BV\Vd(t)-\BV W,
\]
and, therefore,
\begin{multline}
\label{eq:Fext2}
\parderiv{\bar F_0}t=\div_{\BV Z}\int(\EE\BV\Vd(s)-\cL\Phi(s,\BV\Vd))
F^{ext}_1\dif\BV X\dif\BV V\dif\BV\Xd\dif\BV\Vd=\\=\div_{\BV Z}\left(
\int\EE\BV\Vd(s)F^{ext}_1\dif\BV X\dif\BV V\dif\BV\Xd\dif\BV\Vd-\int
\Phi(s,\BV\Vd)\cL^*F^{ext}_1\dif\BV X\dif\BV V\dif\BV\Xd\dif\BV\Vd
\right)=\\=\div_{\BV Z}\left(\int\EE\BV\Vd(s)F^{ext}_1 \dif\BV X\dif
\BV V\dif\BV\Xd\dif\BV\Vd+\int\Phi(s,\BV\Vd)(\BV\Vd\cdot\nabla_{\BV Z}
F^{ext}_0)\dif\BV X\dif\BV V\dif\BV\Xd\dif\BV\Vd \right),
\end{multline}
where in the last line we used~\eqref{eq:e1}, and the parameter $s>0$
is for now unspecified. For the first term in the last line
of~\eqref{eq:Fext2}, observe that
\begin{multline*}
\int\EE\BV\Vd(s)F^{ext}_1 \dif\BV X\dif\BV V\dif\BV\Xd\dif\BV\Vd=
\int\EE\BV\Vd(s)(F^{ext}_0+F^{ext}_1)\dif\BV X\dif\BV V\dif\BV\Xd
\dif\BV\Vd-\\-\int\EE\BV\Vd(s)F^{ext}_0 \dif\BV X\dif\BV V\dif\BV
\Xd\dif\BV\Vd=\int\EE\BV\Vd(s)(F^{ext}_0+F^{ext}_1)\dif\BV X\dif\BV
V\dif\BV\Xd\dif\BV\Vd,
\end{multline*}
where the last identity is due to~\eqref{eq:e0} and the centering
condition. Now observe that $F^{ext}_0+F^{ext}_1$ is a probability
density by itself (as it satisfies the normalization requirement),
which can be viewed as a perturbation from the invariant state
$F^{ext}_0$. Then, as $s\to\infty$, we can assume that a statistical
ensemble of~\eqref{eq:dyn_sys_real}+\eqref{eq:dyn_sys_diff}, initially
distributed according to $F^{ext}_0+F^{ext}_1$, becomes distributed
according to $F^{ext}_0$:
\[
\lim_{s\to\infty}\int\EE\BV\Vd(s)(F^{ext}_0+F^{ext}_1)\dif\BV X\dif\BV
V\dif\BV\Xd\dif\BV\Vd=\int\BV\Vd F^{ext}_0\dif\BV X\dif\BV V\dif\BV
\Xd\dif\BV\Vd=0.
\]
Thus, we arrive at
\begin{equation}
\label{eq:Fext3}
\parderiv{\bar F_0}t=\div_{\BV Z}\left(\lim_{s\to\infty}\int\Phi(s,
\BV\Vd)(\BV\Vd \cdot\nabla_{\BV Z}F^{ext}_0)\dif\BV X\dif\BV V\dif
\BV\Xd\dif\BV\Vd\right),
\end{equation}
as long as the limit in the right-hand side is finite. For this,
observe that the expression in brackets in the right-hand side
of~\eqref{eq:Fext3} can be written as the integral of a time
autocorrelation function:
\begin{multline*}
\lim_{s\to\infty}\int\Phi(s,\BV\Vd)(\BV\Vd\cdot\nabla_{\BV Z}
F^{ext}_0)\dif\BV X\dif\BV V\dif\BV\Xd\dif\BV\Vd=\\=
\lim_{r\to\infty}\int_0^r\dif s\int\EE\BV\Vd(s)(\BV\Vd\cdot
\nabla_{\BV Z}F^{ext}_0)\dif\BV X\dif \BV V\dif\BV\Xd\dif\BV\Vd.
\end{multline*}
The finiteness of the above expression requires that the time
autocorrelation function decays to zero sufficiently rapidly; the
decay to zero can be shown via the mixing and centering conditions
for~\eqref{eq:dyn_sys_real}+\eqref{eq:dyn_sys_diff}:
\begin{multline*}
\lim_{s\to\infty}\int\EE\BV\Vd(s)(\BV\Vd\cdot \nabla_{\BV Z}
F^{ext}_0)\dif\BV X\dif \BV V\dif\BV\Xd\dif\BV\Vd=\\=\int\BV\Vd
F^{ext}_0\dif\BV X\dif \BV V\dif\BV\Xd\dif\BV\Vd\cdot\int\BV\Vd
\cdot \nabla_{\BV Z}F^{ext}_0\dif\BV X\dif\BV V\dif\BV\Xd\dif\BV
\Vd=0.
\end{multline*}
Thus, we finally arrive at
\begin{equation}
\label{eq:Fext4}
\parderiv{\bar F_0}t=\div_{\BV Z}\left(\int_0^\infty\dif s\int\EE\BV
\Vd(s)(\BV\Vd \cdot\nabla_{\BV Z}F^{ext}_0)\dif\BV X\dif\BV V\dif\BV
\Xd\dif\BV\Vd\right).
\end{equation}

\subsection{Simplified relations for the long-term behavior of a gas}

Above, the expectation $\EE\BV\Vd(s)$ is generally a function of the
initial conditions $\BV X$, $\BV V$, $\BV\Xd$ and $\BV\Vd$ for the
joint process \eqref{eq:dyn_sys_real}+\eqref{eq:dyn_sys_diff}.
Apparently, there is no tractable way to compute $\EE\BV\Vd(s)$
exactly, so we need a simple enough approximation for it. In what
follows, we will introduce various simplifications into the right-hand
side of~\eqref{eq:Fext4}, based on the assumption that the statistical
properties of the gas molecule behavior are sufficiently similar, in
certain aspects, to those of some exactly solvable models of the
Brownian motion.

We start by decomposing the invariant probability measure $\im$ of
\eqref{eq:dyn_sys_real}+\eqref{eq:dyn_sys_diff} into the product of
the marginal measure $\im_1(\BV\Vd)$ and the conditional measure
$\im_2(\BV X,\BV V,\BV\Xd|\BV\Vd)$, such that, for a $\im$-measurable
function $\psi$,
\[
\int\psi\dif\im=\int\psi\dif\im_2\dif\im_1.
\]
With the decomposition above, first, we assume that $\EE\BV\Vd(s)$ is
independent of the rest of the integrand above in~\eqref{eq:Fext4} in
$\im_2$, that is,
\begin{multline}
\label{eq:eev_independent}
\int\EE\BV\Vd(s)(\BV\Vd \cdot\nabla_{\BV Z}F^{ext}_0)\dif\BV X\dif
\BV V\dif\BV\Xd\dif\BV\Vd=\\=\int\left(\int\EE\BV\Vd(s)\dif\im_2
\right)(\BV\Vd\cdot\nabla_{\BV Z}F^{ext}_0)\dif\BV X\dif\BV V\dif\BV
\Xd\dif\BV\Vd.
\end{multline}
Clearly, the approximation above is valid for sufficiently small $s$
(where $\EE\BV\Vd(s)\approx\BV\Vd$), and for sufficiently large $s$
due to strong mixing. We assume that this approximation is good enough
for the whole range of $s$.

Next, we assume that the conditional expectation of $\BV\Vd(s)$ in
$\im_2$ behaves as the one for a Gaussian random process (such as, for
example, the Ornstein-Uhlenbeck process \cite{OrnUhl}):
\begin{equation}
\label{eq:gaussian_process}
\int\EE\BV\Vd(s)\dif\im_2=\BM C(s)\BM C^{-1}(0)\BV\Vd, \qquad\BM
C(s)=\int\EE\BV\Vd(s)\otimes\BV\Vd\dif\im.
\end{equation}
With~\eqref{eq:eev_independent} and~\eqref{eq:gaussian_process},
\eqref{eq:Fext4} becomes
\begin{equation}
\label{eq:Fext5}
\parderiv{\bar F_0}t=\div_{\BV Z}\left[\left(\int_0^\infty\BM C(s)\dif
  s\right)\BM C^{-1}(0)\int(\BV\Vd\otimes\BV\Vd)\nabla_{\BV Z}
  F^{ext}_0\dif\BV X\dif\BV V\dif\BV\Xd\dif\BV\Vd\right].
\end{equation}
The second integral in the right-hand side of~\eqref{eq:Fext5} can be
expressed as
\begin{multline*}
\int(\BV\Vd\otimes\BV\Vd)\nabla_{\BV Z} F^{ext}_0\dif\BV X\dif\BV V
\dif\BV\Xd\dif\BV\Vd=\div_{\BV Z}\int(\BV\Vd\otimes\BV\Vd) F^{ext}_0
\dif\BV X\dif\BV V\dif\BV\Xd\dif\BV\Vd=\\=\div_{\BV Z} \left( \bar
F_0\int\BV\Vd\otimes\BV\Vd\dif\im\right)=\div_{\BV Z} (\bar F_0 \BM
C(0)),
\end{multline*}
which results in
\begin{equation}
\parderiv{\bar F_0}t=\div_{\BV Z}\left[\left(\int_0^\infty\BM C(s)
  \dif s\right)\BM C^{-1}(0)\div_{\BV Z}(\bar F_0\BM C(0))\right].
\end{equation}
Above, $\BM C(s)$ is a $KN\times KN$ matrix (and so is its inverse),
which is, again, practically intractable. Thus, we are further going
to assume that $\BM C(0)$ and the integral of $\BM C(s)$ are diagonal,
\[
\BM C(0)=\theta_{\BV\Vd}\BM I,\qquad\int_0^\infty\BM C(s)\dif s=
\eta_{\BV\Vd}\BM I,
\]
where $\BM I$ is a $KN\times KN$ identity matrix. This assumption
ultimately leads to the following closed form of the equation for
$\bar F_0$:
\begin{equation}
\label{eq:Fext6}
\parderiv{\bar F_0}t=\div_{\BV Z}\left(\frac{\eta_{\BV\Vd}}{
  \theta_{\BV\Vd}} \nabla_{\BV Z} (\theta_{\BV\Vd}\bar F_0)\right).
\end{equation}
Above,
\begin{subequations}
\label{eq:theta_eta}
\begin{equation}
\theta_{\BV\Vd}=\frac 1{KN}\int\|\BV\Vd\|^2\dif\im=\int w^2\dif\im,
\end{equation}
\begin{equation}
\eta_{\BV\Vd}=\frac 1{KN}\int_0^\infty\dif s\int\BV\Vd(s)\cdot\BV
\Vd(0)\dif\im=\int_0^\infty\dif s\int w(s)w(0)\dif\im,
\end{equation}
\end{subequations}
where $w(t)$ is an arbitrary component of $\BV\Vd(t)$. The latter
identity is due to the assumption that all components of $\BV\Vd(t)$
are distributed identically at statistical equilibrium.

From~\eqref{eq:Fext6}, the corresponding closed approximate equation
for $\BV Z(t)$ in the homogenization time limit is given by the
following It\^o stochastic differential equation \cite{Ito,Ito2}:
\begin{equation}
\label{eq:homogenization_X}
\dif\bar{\BV Z}=\theta_{\BV\Vd}\nabla_{\BV Z}\left(\frac{\eta_{\BV\Vd}}{
  \theta_{\BV\Vd}}\right)\dif t+\sqrt{2\eta_{\BV\Vd}}\dif\BV\cW(t),
\end{equation}
where $\BV\cW(t)$ is a $KN$-dimensional Wiener process
\cite{GikSko,Oks}.

Observe that the simplified closed dynamics for $\BV Z(t)$ in the
homogenization time limit are controlled by two statistical
quantities, $\theta_{\BV\Vd}$ and $\eta_{\BV\Vd}$, which, in turn, are
given by~\eqref{eq:theta_eta}. For the practical computation, we need
to connect $\theta_{\BV\Vd}$ and $\eta_{\BV\Vd}$ to the appropriate
statistical properties of~\eqref{eq:dyn_sys_real}
and~\eqref{eq:dyn_sys_alt}.

For a simplified estimate of $\theta_{\BV\Vd}$, let us neglect the
influence of the potential energy in~\eqref{eq:dyn_sys_real}
(observing that the potential $H(\BV X)$ has a very short range,
compared to the average distance between the molecules), and assume
that the velocity processes $\BV V(t)$ for both the realistic gas
system in~\eqref{eq:dyn_sys_real} and the random jump model
in~\eqref{eq:dyn_sys_alt} are uniformly distributed on the
$((K-1)N-1)$-dimensional constant momentum/energy sphere of the same
radius. Therefore, $\theta_{\BV\Vd}$ is the average square distance
between two points on the same $((K-1)N-1)$-dimensional sphere, and
can be related to the radius of the sphere via geometric
arguments. Namely, let one of the points be at a pole of the sphere of
radius 1, and let the other slide along a meridian.  Let $x$ be the
distance between the first point, and the projection of the second
point onto the axis (such that $0\leq x\leq 2$). Then, the distance
between the points is
\[
\text{distance}=\sqrt{x^2+1-(x-1)^2}=\sqrt{2x}.
\]
We also have to observe that the second point is uniformly distributed
over a sphere of codimension 1 and radius
$\sqrt{1-(x-1)^2}=\sqrt{2x-x^2}$, whose relative (to the full
$((K-1)N-1)$-dimensional sphere of radius 1) surface area is given by
\[
S_{(K-1)N-1}(1)=\frac{2\pi^{(K-1)N/2}}{\Gamma((K-1)N/2)},
\]
\[
S_{(K-1)N-2}(\sqrt{2x-x^2})=\frac{2\pi^{((K-1)N-1)/2}}{\Gamma(((K-1)
  N-1)/2)}(2x-x^2)^{((K-1)N-2)/2},
\]
\[
\frac{S_{(K-1)N-2}}{S_{(K-1)N-1}}\approx\sqrt{\frac{(K-1)N-1}{2\pi}}
(2x-x^2)^{((K-1)N-2)/2},
\]
where the approximation is valid in the limit as $KN\to\infty$. To
compute the average square distance, we, therefore, need to integrate
the squared distance against the ratio above, which results in
\begin{multline*}
\int_0^22x\frac{S_{(K-1)N-2}}{S_{(K-1)N-1}}\dif x\approx 2\sqrt{\frac{
    (K-1)N-1}{2\pi}} \int_0^2 (2x-x^2)^{((K-1)N-2)/2}x\dif x=\\=2
\sqrt{\frac{(K-1)N-1}{2\pi}} \int_{-1}^1 (1-y^2)^{((K-1)N-2)/2}\dif y
\approx\\\approx 2\sqrt{\frac{(K-1)N-1}{2\pi}} \sqrt{\frac 2{(K-1)
    N-2}} \int_{-\infty}^\infty e^{-z^2}\dif z\approx 2,
\end{multline*}
as $KN\to\infty$. This means that if the squared radius of the
constant energy sphere of the random jump model
in~\eqref{eq:dyn_sys_alt} (or, equivalently, the realistic gas model
in~\eqref{eq:dyn_sys_real}) is given by $\theta_0$, then
\begin{equation}
\label{eq:theta_d_0}
\theta_{\BV\Vd}=2\theta_0.
\end{equation}
The key observation here is that $\theta_{\BV\Vd}$ is a constant
multiple of $\theta_0$, regardless of what the value of the constant
factor actually is.  As a result, the equation for $\bar F_0$
in~\eqref{eq:Fext6} becomes
\begin{equation}
\parderiv{\bar F_0}t=\div_{\BV Z}\left(\frac{\eta_{\BV\Vd}}{\theta_0}
\nabla_{\BV Z} (\theta_0\bar F_0)\right).
\end{equation}
For $\eta_{\BV\Vd}$, the situation is generally more complicated, as
the time evolution of $\EE\BV\Vd(t)$ is governed by a different
operator than those of either $\EE\BV V(t)$
from~\eqref{eq:dyn_sys_alt} or $\BV V(t)$
from~\eqref{eq:dyn_sys_real}. Thus, the decay of the velocity time
autocorrelation functions in~\eqref{eq:dyn_sys_real},
\eqref{eq:dyn_sys_alt} and~\eqref{eq:dyn_sys_diff} can be quite
different, and relating them to each other in a detailed fashion
appears to be a rather complicated task. On the other hand, at this
point we already found in~\eqref{eq:theta_d_0} that $\theta_{\BV\Vd}$
and $\theta_0$ are constant multiples of each other. Therefore, here
we make a ``physicist's common sense'' assumption that $\eta_{\BV\Vd}$
and the corresponding integral of the velocity time autocorrelation
function of~\eqref{eq:dyn_sys_alt}, which we denote as $\eta_0$, are
also related by an empirical constant $\alpha$, which can be estimated
from the observations:
\begin{equation}
\eta_{\BV\Vd}=\alpha\eta_0.
\end{equation}
As a result, we arrive at
\begin{equation}
\label{eq:Fext7}
\parderiv{\bar F_0}t=\div_{\BV Z}\left(\frac{\alpha\eta_0}{\theta_0}
\nabla_{\BV Z} (\theta_0\bar F_0)\right).
\end{equation}

\section{The diffusive Boltzmann equation}
\label{sec:diffusive_boltzmann}

Observe that the process $\BV Z(t)$ back
in~\eqref{eq:dyn_sys_rescaled} was chosen as the negative of the
difference coordinate process $\BV\Xd(t)$, which means that adding
$\BV Z(t)$ to the solution of the random jump system
in~\eqref{eq:dyn_sys_alt} yields the solution of the realistic gas
system in~\eqref{eq:dyn_sys_real}. Therefore, here we propose to
correct the random jump system in~\eqref{eq:dyn_sys_alt} towards the
realistic gas system in~\eqref{eq:dyn_sys_real} by adding the It\^o
diffusion process in~\eqref{eq:homogenization_X}
to~\eqref{eq:dyn_sys_alt}, obtaining
\begin{equation}
\label{eq:new_system}
\dif\BV X=\left(\BV V+\theta_0\nabla_{\BV X}\left(\frac{\alpha\eta_0}{
  \theta_0}\right)\right)\dif t+\sqrt{2\alpha\eta_0}\dif\BV\cW(t),
\qquad\dif\BV V=\dif\BV\cN(t).
\end{equation}
This is a somewhat different Feller process, whose forward Kolmogorov
equation is given by
\begin{equation}
\label{eq:new_gen_kolm}
\parderiv Ft+\BV V\cdot\nabla_{\BV X}F=\int G^B(\BV X;\BV V'|\BV V)
\big[F(\BV V')-F(\BV V)\big]\dif\BV V'+\div_{\BV X}\left(\frac{\alpha
  \eta_0}{\theta_0}\nabla_{\BV X}(\theta_0 F)\right).
\end{equation}
Following the same steps as in Section~\ref{sec:random_jumps}, from
the Kolmogorov equation in~\eqref{eq:new_gen_kolm} we obtain the new
diffusive Boltzmann equation
\begin{equation}
\label{eq:diff_boltzmann_2}
\parderiv ft+\BV v\cdot\nabla_{\BV x}f=\coll(f)+\div_{\BV x}\left(
\frac{\alpha\eta_0}{\theta_0}\nabla_{\BV x}(\theta_0 f)\right).
\end{equation}
Observe that~\eqref{eq:new_gen_kolm} has the same microcanonical Gibbs
state as~\eqref{eq:gen_kolm}, while its corresponding diffusive
Boltzmann equation in~\eqref{eq:diff_boltzmann_2} has the same
Gaussian steady solution as the original Boltzmann equation
in~\eqref{eq:boltzmann}. Thus, we are going to apply the same
approximations between the microcanonical and canonical Gibbs states
used above in Section~\ref{sec:random_jumps}.

Here observe that~$\theta_0$ and~$\eta_0$ are the corresponding
velocity variance and integrated time autocorrelation function of the
multimolecule random jump system in~\eqref{eq:dyn_sys_alt} at
statistical equilibrium. However, if we are solving the diffusive
Boltzmann equation in~\eqref{eq:diff_boltzmann_2} in a practical
situation, its solution $f$ is not necessarily at statistical
equilibrium. For the lack of any better approximation, we will have to
assume that $f$ yields sufficiently good approximations of $\theta_0$
and $\eta_0$ in practice.

Let us introduce a concise notation of an arbitrary statistical moment
of the single-molecule distribution function~$f$ from the diffusive
Boltzmann equation in~\eqref{eq:diff_boltzmann_2}. Let~$g(\BV v)$ be
an integrable (with respect to~$f$) function of~$\BV v$. We then
denote
\begin{equation}
\langle g\rangle_f(t,\BV x) = \int gf\dif\BV v,\qquad \langle
g\rangle_{\coll(f)}(t,\BV x) = \int g\,\coll(f)\dif\BV v,
\end{equation}
where~$\coll(f)$ is the Boltzmann collision operator. Now, we write
the density~$\rho$, flow velocity~$\BV u$, energy~$E$,
temperature~$\theta$ and pressure~$p$ as
\begin{equation}
\label{eq:rho_u_theta}
\rho=\langle 1\rangle_f,\quad\rho\BV u=\langle\BV v\rangle_f,\quad
2\rho E=\langle\|\BV v\|^2 \rangle_f, \quad N\theta=2E-\|\BV u\|^2,
\quad p=\rho\theta.
\end{equation}
We also introduce the mass diffusion coefficient $D$ and its
empirically $\alpha$-scaled version $D_\alpha$ as
\begin{equation}
\label{eq:D}
D=\int_0^\infty\langle (v(s)-u)(v(0)-u)\rangle_f\dif s,\qquad
D_\alpha=\alpha D,
\end{equation}
where $v(s)$ is the time series of an arbitrary component of $\BV
V(s)$ of the random jump system~\eqref{eq:dyn_sys_alt}, and $u$ is the
corresponding mean velocity component.

So, our approximations for $\theta_0$ and $\eta_0$ here are quite
straightforward:
\begin{equation}
\theta_0\approx\theta,\qquad\alpha\rho\eta_0\approx D_\alpha.
\end{equation}
As a result, we write the diffusive Boltzmann equation
in~\eqref{eq:diff_boltzmann_2} as
\begin{equation}
\label{eq:diff_boltzmann}
\parderiv ft+\BV v\cdot\nabla_{\BV x}f=\coll(f)+\div_{\BV x}\left(
\frac{D_\alpha}p\nabla_{\BV x}(\theta f)\right).
\end{equation}

\section{The diffusive fluid dynamics equations}
\label{sec:fluid_dynamics}

Here we integrate the diffusive Boltzmann equation against different
powers of the velocity~$\BV v$, obtaining the equations for different
velocity moments of the distribution density~$f$. This is a standard
procedure, which, for the usual Boltzmann equation
in~\eqref{eq:boltzmann}, leads to the conventional equations of gas
dynamics, such as the Euler and Grad equations
\cite{Bat,Lev,Gols,Gra}.

Integrating the diffusive Boltzmann equation
in~\eqref{eq:diff_boltzmann} against a moment~$g(\BV v)$ in~$\BV v$,
and using the notations from the previous section, we obtain
\begin{equation}
\label{eq:moment_equation}
\parderiv{\langle g\rangle_f}t+\div\langle g\BV v\rangle_f=\langle g
\rangle_{\coll(f)}+\div\left(\frac{D_\alpha}p\nabla(\theta\langle g
\rangle_f)\right),
\end{equation}
where we drop the subscript ``$\BV x$'' from the differentiation
operators, since the $\BV v$-variable is no longer in the equation. It
is interesting that the additional term in the right-hand side can be
separated into the diffusion and transport terms as
\begin{equation}
\div\left(\frac{D_\alpha}p\nabla(\theta\langle g\rangle_f)\right)=
\div\left(\frac{D_\alpha}p\langle g\rangle_f\nabla\theta\right)+
\div\left(\frac{D_\alpha} \rho\nabla\langle g\rangle_f\right),
\end{equation}
which, together, constitute a simple linearized thermophoretic
transport-diffusion process with the Soret coefficient \cite{DuhBra}
set to~$\theta^{-1}$.  In the earlier works on the extended fluid
dynamics \cite{Brenner,Brenner2,Brenner3,DadRee}, the transport terms
due to the temperature gradient are not present. Surprisingly, in
\cite{Durst,Durst2} the temperature gradient transport terms are
present, however, they appear to be of a form different from the one
above.

Observe that the moment equation above in~\eqref{eq:moment_equation}
is not automatically closed with respect to~$\langle g\rangle_f$, as
there is a higher order term~$\langle g\BV v\rangle_f$ present.
Different closures of the moment equations lead to different
hierarchies of the corresponding fluid dynamics equations. Below, we
re-derive the Euler, Navier-Stokes \cite{Bat,Gols,Lev}, Grad
\cite{Gra,Gra2} and regularized Grad \cite{Stru,StruTor,TorStru}
equations from the diffusive Boltzmann equation in a standard way.

\subsection{The diffusive Euler equations}

It is well known (see, for example, \cite{Cer,Gols,Gra,Lev}) that the
collision moments of the density, momentum and energy are zeros, due
to the mass, momentum, and energy conservation during the collisions
of the molecules:
\begin{equation}
\langle 1\rangle_{\coll(f)}=0,\qquad \langle\BV v \rangle_{\coll(f)}
=\BV 0,\qquad \langle\|\BV v\|^2 \rangle_{\coll(f)}=0.
\end{equation}
Thus, the transport equations for these moments read
\begin{subequations}
\label{eq:moment_equations_euler}
\begin{equation}
\parderiv\rho t+\div\langle\BV v\rangle_f=\div\left(\frac{D_\alpha}p
\nabla p\right),
\end{equation}
\begin{equation}
\parderiv{(\rho\BV u)}t+\div\langle\BV v\otimes\BV v \rangle_f
=\div\left(\frac{D_\alpha}p\Big(\nabla\otimes(p\BV u)\Big)\right),
\end{equation}
\begin{equation}
\parderiv{(\rho E)}t+\frac 12\div\langle \|\BV v\|^2\BV v\rangle_f
=\div\left(\frac{D_\alpha}p\nabla(pE)\right),
\end{equation}
\end{subequations}
where the divergence of a tensor contracts over its first index. Let
the $N\times N$-dimensional temperature matrix $\BM T$ and the
$N$-dimensional heat flux vector $\BV q$ via the relations
\begin{equation}
\label{eq:Tq2}
\rho\BM T=\langle(\BV v-\BV u)\otimes(\BV v-\BV u)\rangle_f, \qquad
\rho\BV q=\frac 12\langle\|\BV v-\BV u\|^2(\BV v-\BV u)\rangle_f.
\end{equation}
Note that here we choose to normalize the heat flux~$\BV q$ by the
density~$\rho$ for convenience. This is unlike the conventional
notation in, for example, \cite{Gra}, where the heat flux is given by
the corresponding unnormalized moment. However, here we prefer the
notations in~\eqref{eq:Tq2} due to the fact that they are analogous to
the velocity~$\BV u$ and temperature~$\theta$, and thus are more
suitable for defining the boundary conditions for gas flows. Indeed,
observe that it is the velocity~$\BV u$ and temperature~$\theta$ which
are typically specified at the boundaries, as opposed to the
momentum~$\rho\BV u$ and pressure~$p$.

One then can verify directly that~$\BM T$ and~$\BV q$ satisfy the
relations
\begin{equation}
\label{eq:Tq}
\rho(\BV u\otimes\BV u+\BM T)=\langle\BV v\otimes\BV v\rangle_f,
\qquad\rho\left(E\BV u+\BM T\BV u+\BV q\right)=\frac 12\langle\|\BV
v\|^2\BV v\rangle_f.
\end{equation}
Clearly, the equations in~\eqref{eq:moment_equations_euler} above are
closed with respect to $\rho$, $\BV u$ and $\theta$, but not with
respect to $\BM T$ or $\BV q$. The Euler closure
for~\eqref{eq:moment_equations_euler} is achieved under the assumption
that the molecule velocity distribution $f$ is equal to the
Maxwell-Boltzmann statistical equilibrium \cite{Lev,Gols}, given in
the form
\begin{equation}
\label{eq:MB_state}
f_{MB}=\frac\rho{(2\pi\theta)^{N/2}}\exp\left(-\frac{\|\BV v-\BV u\|^2
}{2\theta}\right),
\end{equation}
which sets the temperature matrix to $\BM T=\theta\BM I$, and the heat
flux $\BV q$ to zero. As a result, the moment transport equations
in~\eqref{eq:moment_equations_euler} become
\begin{subequations}
\label{eq:Euler_equations}
\begin{equation}
\parderiv\rho t+\div(\rho\BV u)=\div\left(\frac{D_\alpha}p \nabla
p\right),
\end{equation}
\begin{equation}
\parderiv{(\rho\BV u)}t+\div\left(\rho\BV u\otimes\BV u\right)+\nabla
p=\div\left(\frac{D_\alpha}p\Big(\nabla\otimes(p\BV u)\Big) \right),
\end{equation}
\begin{equation}
\parderiv{}t\left(\rho E \right)+\div\left(\rho\left(E+\theta
\right)\BV u\right)=\div\left(\frac{D_\alpha}p\nabla \left( pE\right)
\right).
\end{equation}
\end{subequations}
The equations in~\eqref{eq:Euler_equations} above are the diffusive
analogs of the well-known Euler equations \cite{Bat,Gols,Lev}. Observe
that the new diffusion term from~\eqref{eq:diff_boltzmann} manifests
itself in all moment equations.

Above, the $\BV x$-differentiation is formally done for each degree of
freedom of a gas molecule (that is, 3 translational degrees, and
$(N-3)$ rotational degrees). However, in practical situations it is
assumed that the moment averages are distributed uniformly along the
rotational degrees (that is, the orientation angles of molecules), and
thus only the translational $\BV x$-differentiations are often taken
into account. Also, it is usually assumed that the rotational
components of the momentum~$\langle\BV v\rangle_f$ are zero, and thus
the equations above are entirely closed with respect to the
density~$\rho$, the temperature~$\theta$, and the translational
components of the velocity~$\BV u$ as functions of the translational
coordinates of~$\BV x$.

One can also write the separate equation for the pressure $p$. For
that, observe that
\begin{subequations}
\begin{equation}
\frac 12\parderiv{}t\left(\rho\|\BV u\|^2\right)=\BV
u\cdot\parderiv{}t(\rho\BV u)-\frac 12\|\BV u\|^2\parderiv\rho t,
\end{equation}
\begin{equation}
\frac 12\div\left(\rho\|\BV u\|^2\BV u\right)=\BV u\cdot\div(\rho\BV
u\otimes\BV u)-\frac 12\|\BV u\|^2\div(\rho\BV u),
\end{equation}
\begin{multline}
\frac 12\div\left(\frac{D_\alpha}p\nabla(p\|\BV u\|^2)\right)=\BV
u\cdot\div\left(\frac{D_\alpha}p\Big( \nabla\otimes(p\BV u)\Big)
\right)-\\-\frac 12\|\BV u\|^2\div\left( \frac{D_\alpha}p\nabla
p\right)+D_\alpha\|\nabla\otimes\BV u\|^2,
\end{multline}
\end{subequations}
and thus, subtracting the appropriate multiples of the density and
momentum equations, we obtain
\begin{equation}
\parderiv pt+\div(p\BV u)+(\gamma-1)p\, \div\BV
u=\div\left(\frac{D_\alpha}p\nabla(p\theta)\right)
+(\gamma-1)D_\alpha\|\nabla\otimes\BV u\|^2,
\end{equation}
where we introduced the adiabatic exponent
\begin{equation}
\gamma=1+\frac 2N,
\end{equation}
and $\|\nabla\otimes\BV u\|^2$ is given by
\begin{equation}
\|\nabla\otimes\BV u\|^2=(\nabla\otimes\BV u):(\nabla\otimes\BV u),
\end{equation}
where ``$:$'' denotes the Frobenius product of two matrices. Observe
that the diffusive Euler equations above are of the second order
in~$\BV x$, and thus are more suitable for boundary value problems,
than the conventional Euler equations (which are of the first order
in~$\BV x$).

\subsection{The diffusive Navier-Stokes equations}

The conventional Navier-Stokes equations \cite{Bat,Lev,Gols} are
obtained as an ``upgrade'' from the conventional Euler equations as
follows. First, it is no longer assumed that $\BM T=\theta\BM I$, and
$\BV q=\BV 0$, and the deviator ``stress'' matrix $\BM S$ between $\BM
T$ and $\theta\BM I$ is introduced:
\begin{equation}
\label{eq:S}
\BM S=\BM T-\theta\BM I.
\end{equation}
Note that here the stress $\BM S$ is normalized by the density in the
same way as the heat flux $\BV q$ above.

Second, the translational components of the stress $\BM S$ and heat
flux $\BV q$ are approximated via the Newton and Fourier laws, under
the assumption that the dynamics of the stress and heat flux manifests
itself on a much faster time scale and is thus ``slaved'' to the
slower dynamics of the density, momentum and energy. The Newton law
for the stress $\BM S$ and the Fourier law for the heat flux $\BV q$
are given, respectively, by
\begin{subequations}
\label{eq:Sq_NS}
\begin{equation}
\rho\BM S_{Newton}=-\mu\left(\nabla\otimes\BV u+(\nabla\otimes\BV u)^T
+(1-\gamma)(\div\BV u)\BM I\right),
\end{equation}
\begin{equation}
\rho\BV q_{Fourier}=-\frac\gamma{\gamma-1}\frac\mu\Pran\nabla\theta,
\end{equation}
\end{subequations}
where~$\mu$ is the dynamic viscosity and~$\Pran$ is the Prandtl
number. For the rotational components of $\BM S$ and $\BV q$ (just
like for the rotational components of $\BV u$ above), it is assumed
that they are uniformly distributed in the corresponding rotational
coordinates, so that their divergences in rotational coordinates are
zero. Thus, the rotational components of $\BM S$ and $\BV q$ are
decoupled from the translational motion and no approximation for them
is necessary.

Here we obtain the diffusive Navier-Stokes equations from the
diffusive Euler equations in~\eqref{eq:Euler_equations} in exactly the
same manner, by substituting the Newton and Fourier laws
in~\eqref{eq:Sq_NS} into the moment expressions in~\eqref{eq:Tq}
and~\eqref{eq:S}:
\begin{subequations}
\label{eq:Navier-Stokes}
\begin{equation}
\parderiv\rho t+\div(\rho\BV u)=\div\left(\frac{D_\alpha}p\nabla p
\right),
\end{equation}
\begin{multline}
\parderiv{(\rho\BV u)}t+\div(\rho\BV u\otimes\BV u)+\nabla p
=\div\left(\frac{D_\alpha}p\Big(\nabla\otimes(p\BV u)\Big)\right)
+\\+\div\left(\mu\left(\nabla\otimes\BV u+(\nabla\otimes \BV
u)^T\right)\right)+(1-\gamma)\nabla(\mu\,\div\BV u),
\end{multline}
\begin{multline}
\parderiv pt+\div(p\BV u)+(\gamma-1)p\,\div\BV u=\div\left(\frac{
  D_\alpha}p\nabla(p\theta)\right)+\frac\gamma\Pran\div(\mu\nabla
\theta)+\\+(\gamma-1)\left((D_\alpha+\mu) \|\nabla\otimes\BV u\|^2
+\mu\left((\nabla\otimes\BV u): (\nabla\otimes\BV u)^T+(1-\gamma)
(\div\BV u)^2\right)\right).
\end{multline}
\end{subequations}
The diffusive Navier-Stokes equations above are somewhat similar to
those in the works on the extended fluid
dynamics \cite{Brenner,Brenner2,Brenner3,DadRee,Durst,Durst2}.

\subsection{The diffusive Grad equations}

For the diffusive Grad equations, we augment the existing transport
equations in~\eqref{eq:moment_equations_euler} with the new transport
equations for the stress and heat flux moments, given by
\begin{subequations}
\label{eq:moment_equations_grad}
\begin{equation}
\parderiv{\langle\BV v\otimes\BV v\rangle_f}t+\div\langle\BV v\otimes
\BV v\otimes\BV v\rangle_f=\langle\BV v\otimes\BV v\rangle_{\coll(f)}+
\div\left(\frac{D_\alpha}{\rho\theta}\Big(\nabla\otimes(\theta\langle
\BV v\otimes\BV v\rangle_f)\Big)\right),
\end{equation}
\begin{multline}
\frac 12\parderiv{\langle\|\BV v\|^2\BV v\rangle_f}t+\frac 12\div
\langle\|\BV v\|^2\BV v\otimes\BV v\rangle_f=\\=\frac 12 \langle\|\BV
v\|^2\BV v\rangle_{\coll(f)}+\frac 12\div\left(\frac{D_\alpha}{\rho
  \theta}\Big( \nabla\otimes(\theta\langle\|\BV v\|^2\BV
v\rangle_f)\Big)\right).
\end{multline}
\end{subequations}
Observe that the collision terms are nonzero for the stress~$\BM S$
and heat flux~$\BV q$ (as opposed to the density, momentum and
energy). Also, the new equations include the unknown higher-order
moments $\langle\BV v\otimes\BV v\otimes\BV v\rangle_f$ and
$\langle\|\BV v\|^2\BV v\otimes\BV v\rangle_f$. For these higher-order
moments, we introduce the corresponding centered moments
\begin{equation}
\BM Q=\frac 1\rho\langle(\BV v-\BV u)\otimes(\BV v-\BV u)\otimes(\BV v
-\BV u)\rangle_f,\qquad\BM R=\frac 1{2\rho}\langle\|\BV v-\BV
u\|^2(\BV v-\BV u)\otimes(\BV v-\BV u)\rangle_f,
\end{equation}
with~$\BM Q$ being the $N\times N\times N$ 3-rank tensor, and~$\BM R$
being the $N\times N$ matrix. One can verify that~$\BM Q$ and~$\BM R$
satisfy the identities
\begin{subequations}
\begin{equation}
\frac 1\rho\langle\BV v\otimes\BV v\otimes\BV v\rangle_f=\BV u\otimes
\BV u\otimes\BV u+\BM T\otimes\BV u+(\BM T\otimes\BV u)^T+(\BM T
\otimes \BV u)^{TT}+\BM Q,
\end{equation}
\begin{equation}
\frac 1{2\rho}\langle\|\BV v\|^2\BV v\otimes\BV v\rangle_f=E\BV
u\otimes\BV u+\frac 12\|\BV u\|^2\BM T+(\BM T\BV u+\BV q) \otimes\BV
u+\BV u\otimes(\BM T\BV u+\BV q)+\BM Q\BV u+\BM R,
\end{equation}
\end{subequations}
where ``$TT$'' denotes the double transposition of a 3-rank tensor.
In order to approximate the higher-order centered moments~$\BM Q$
and~$\BM R$, we use the Grad distribution \cite{Gra}
\begin{equation}
f_{Grad}=f_{MB}\left(1+\frac 1{2\theta^2}(\BV v-\BV u)^T\BM S(\BV v-
\BV u)+\frac 1 {\theta^2}\left(\frac{\gamma-1}{2\gamma\theta}
\left\|\BV v-\BV u\right\|^2-1\right)\BV q\cdot(\BV v-\BV u)\right),
\end{equation}
which is chosen so as to satisfy the prescribed~$\rho$, $\BV u$,
$\theta$, $\BM S$ and~$\BV q$. For the derivation of~$f_{Grad}$ for a
polyatomic gas molecule, see, for example, \cite{Mal}. For the
higher-order moments~$\BM Q$ and~$\BM R$ we use their Grad
approximations, provided by~$f_{Grad}$:
\begin{subequations}
\begin{equation}
\BM Q_{Grad}=\frac{\gamma-1}\gamma\left(\BM I\otimes\BV q+(\BM I
\otimes\BV q)^T+(\BM I \otimes\BV q)^{TT}\right),
\end{equation}
\begin{equation}
\BM R_{Grad}=\frac\gamma{\gamma-1}\theta\BM T+\theta\BM S.
\end{equation}
\end{subequations}
Substituting the approximations above into the moment relations, we
obtain
\begin{subequations}
\label{eq:higher_moments_Grad}
\begin{multline}
\frac 1\rho\langle\BV v\otimes\BV v\otimes\BV v\rangle_{f_{Grad}}=\BV
u\otimes\BV u\otimes\BV u+\BM T\otimes\BV u+(\BM T\otimes\BV
u)^T+\\+(\BM T\otimes\BV u)^{TT}+\frac{\gamma-1}\gamma\left(\BM
I\otimes\BV q+(\BM I\otimes\BV q)^T+ (\BM I\otimes\BV q)^{TT}\right),
\end{multline}
\begin{multline}
\frac 1{2\rho}\langle\|\BV v\|^2\BV v\otimes\BV v\rangle_{f_{Grad}}=E
\BV u\otimes\BV u+(E+\theta)\BM T+\theta\BM S+\frac{\gamma-1}\gamma(
\BV q\cdot\BV u)\BM I+\\+\left(\BM T\BV u+\frac{ 2\gamma-1}\gamma\BV
q\right) \otimes\BV u+\BV u\otimes \left(\BM T\BV u+\frac{2\gamma-1}
\gamma\BV q \right).
\end{multline}
\end{subequations}
As in \cite{Gra}, we approximate the moment collision terms via the
linear damping:
\begin{equation}
\label{eq:Grad_collision}
\langle\BV v\otimes\BV v\rangle_{\coll(f)}=-\frac{\rho^2\theta}\mu\BM
S,
\qquad
\frac 12\langle\|\BV v\|^2\BV
v\rangle_{\coll(f)}=-\frac{\rho^2\theta}\mu(\BM S\BV u+\Pran\,\BV q).
\end{equation}
As a result, the diffusive Grad equations
in~\eqref{eq:moment_equations_euler}
and~\eqref{eq:moment_equations_grad} become closed with respect to the
variables~$\rho$, $\BV u$, $\theta$, $\BM S$ and~$\BV q$,
via~\eqref{eq:rho_u_theta},~\eqref{eq:Tq},~\eqref{eq:S},~\eqref{eq:higher_moments_Grad}
and~\eqref{eq:Grad_collision}. Observe that the diffusive Grad
equations in~\eqref{eq:moment_equations_euler}
and~\eqref{eq:moment_equations_grad} are of the second order in~$\BV
x$, and thus are usually well posed for a variety of boundary value
problems (unlike the original Grad equations \cite{Gra,Gra2}).

For the reduction of the phase space to the translational components
of the velocity $\BV u$, stress $\BM S$ and heat flux $\BV q$ above in
the diffusive Grad equations~\eqref{eq:moment_equations_euler}
and~\eqref{eq:moment_equations_grad} one has to assume that the
rotational components of the velocity $\BV u$ heat flux $\BV q$ are
zero, and that the spatial derivatives in the rotational directions
are zero. No assumptions need to be made about the rotational
components of the stress matrix, since they become decoupled from the
translational transport equations. Also, observe that if no rotational
components of the phase space are present (for example, if the gas is
monatomic, and the physical space is fully three-dimensional), then
one of the diagonal stress transport equations becomes redundant, due
to the fact that the trace of $\BM S$ is, by construction, zero.

\subsection{The diffusive regularized Grad equations}

Similar to the Navier-Stokes modification~\eqref{eq:Navier-Stokes} of
the Euler equations~\eqref{eq:Euler_equations} via the Newton and
Fourier laws~\eqref{eq:Sq_NS}, the higher-order regularization of the
Grad equations in~\eqref{eq:moment_equations_euler}
and~\eqref{eq:moment_equations_grad} was suggested
in \cite{Stru,StruTor,TorStru} for a monatomic gas. Here we extend the
regularization of the Grad equations onto the polyatomic case and the
diffusive setting, by following the same approach as
in \cite{Stru,StruTor,TorStru}. We present the regularization formulas
for the polyatomic Grad equations while omitting their derivation, due
to the excessive complexity and lengthiness of the latter.

The diffusive regularized Grad equations for a polyatomic gas are
obtained directly from the diffusive Grad equations
in~\eqref{eq:moment_equations_euler}
and~\eqref{eq:moment_equations_grad} by replacing the Grad
approximations for the third- and fourth-order moments $\BM Q_{Grad}$
and $\BM R_{Grad}$ with the regularized approximations. The
expressions for the third order moment $\BM Q$ and the fourth order
moment $\BM R$ in the regularized Grad equations are given by
\begin{subequations}
\label{eq:Reg_Grad}
\begin{equation}
\BM Q_{Reg.Grad}=\BM Q_{Grad}+\widetilde{\!\BM Q}+\widetilde{\!\BM
  Q}^T+\widetilde{\!\BM Q}^{TT},
\end{equation}
\begin{equation}
\BM R_{Reg.Grad}=\BM R_{Grad}+\widetilde{\!\BM R}+\widetilde{\!\BM R
}^T+\Big(\widetilde R+(1-\gamma)\trace(\widetilde{\!\BM R})\Big)\BM I,
\end{equation}
\end{subequations}
where the corrections $\widetilde{\!\BM Q}$, $\widetilde R$ and
$\widetilde{\!\BM R}$ read
\begin{subequations}
\label{eq:Reg_Grad_2}
\begin{multline}
\widetilde{\!\BM Q}=-\frac 1{\Pran_{\widetilde{\!\BM Q}}}\frac\mu\rho\bigg[
  \nabla\otimes\BM S-\frac{\gamma-1}\gamma\BM I\otimes\div\BM S-\frac
  1{\rho\theta}\left(\BM S\otimes\div(\rho\BM S)-\frac{\gamma-1}
  \gamma\BM I\otimes\BM S\,\div(\rho\BM S)\right)+\\+\frac{\gamma-1}{
    \gamma\theta}\bigg(\BV q\otimes\left(\nabla\otimes\BV u+(\nabla
  \otimes\BV u)^T\right)-\frac{\gamma-1}\gamma\BM I\otimes\left(\nabla
  \otimes\BV u+(\nabla\otimes\BV u)^T+(\div\BV u)\BM I\right)\BV q
  \bigg)\bigg],
\end{multline}
\begin{equation}
\widetilde R=-\frac 2{\Pran_{\widetilde R}}\frac\mu\rho\left[\frac
\gamma\theta\div(\theta\BV q)-\div\BV q+(\gamma-1)\left(\BM S:(\nabla
\otimes\BV u)-\frac 1{\rho\theta}\BV q\cdot\div(\rho\BM S)\right)
\right],
\end{equation}
\begin{multline}
\widetilde{\!\BM R}=-\frac 1{\Pran_{\widetilde{\!\BM R}}}\frac\mu\rho
\bigg[\BM S\left(\nabla\otimes\BV u+(\nabla\otimes\BV u)^T\right)+
  \frac{2\gamma -1}{\gamma\theta}\left(\nabla\otimes(\theta\BV
  q)-\frac 1\rho\BV q\otimes\div(\rho\BM S)\right)-\\-\left(
  (\gamma-1)\div\BV u+\frac{2\gamma-1}{2\theta}\left(\frac
  1\rho\div(\rho\BV q) +\BM S:(\nabla\otimes\BV u)\right)\right)\BM
  S\bigg].
\end{multline}
\end{subequations}
Above, the constants $\Pran_{\widetilde{\!\BM Q}}$, $\Pran_{\widetilde
  R}$ and $\Pran_{\widetilde{\!\BM R}}$ are the third- and
fourth-moment Prandtl numbers, which equal $3/2$, $2/3$ and $7/6$,
respectively, for an ideal monatomic gas \cite{Stru,StruTor,TorStru}.
Unfortunately, it does not seem to be possible to compute the exact
values for $\Pran_{\widetilde{\!\BM Q}}$, $\Pran_{\widetilde R}$ and
$\Pran_{\widetilde{\!\BM R}}$ for a polyatomic gas in general, since
the collision between polyatomic gas molecules is a complex process
which depends on the fine structure of a gas molecule. It might be
possible, however, to measure the values of $\Pran_{\widetilde{\!\BM
    Q}}$, $\Pran_{\widetilde R}$ and $\Pran_{\widetilde{\!\BM R}}$
experimentally for common polyatomic gases. In the current work, we
leave the values of $\Pran_{\widetilde{\!\BM Q}}$, $\Pran_{\widetilde
  R}$ and $\Pran_{\widetilde{\!\BM R}}$ for nitrogen at the same
values as for an ideal monatomic gas, since the experimental Prandtl
number of nitrogen ($\sim 0.69$) is not much different from the one of
an ideal monatomic gas ($2/3$). As we find below, the computational
results do not seem to be affected much by such a crude approximation.

\section{A simple computational test: the Couette flow}
\label{sec:numerics}

The Couette flow is a simplest form of a two-dimensional gas flow
between two infinite moving parallel walls. It is assumed that the gas
``sticks'' to the walls to some extent, such that the velocity~$\BV u$
of the flow assumes different values at the boundaries (due to the
walls moving with different speeds relative to each other). Despite
its simplicity, the Couette flow problem is not well-posed for the
conventional Euler or Grad equations, since both are the first-order
differential equations in the space variable~$\BV x$, and thus become
overdetermined in the case of different Dirichlet boundary conditions
at different walls. Instead, conventionally the Couette flow problem
is solved via the famous Navier-Stokes equations \cite{Bat,Lev,Gols},
which, in our case, can be obtained directly from the diffusive
Navier-Stokes equations in~\eqref{eq:Navier-Stokes} by setting the
scaled mass diffusivity $D_\alpha=0$.  Conventionally, the
Navier-Stokes equations are obtained from the hierarchy of the moment
transport equations by the Chapman-Enskog perturbation
expansion \cite{ChaCow,Gols,Lev}, with the Newton and Fourier laws
used to express the stress and heat flux. Observe that the
Navier-Stokes equations in~\eqref{eq:Navier-Stokes} are of the second
order in the momentum and energy, even if the scaled mass diffusivity
$D_\alpha$ is set to zero. This allows to specify the velocity and
temperature at both walls without making the problem overdetermined.

On the other hand, the diffusive Boltzmann equation
in~\eqref{eq:diff_boltzmann} is already of the second order in~$\BV
x$, and so become all its moment equations, including the diffusive
Euler~\eqref{eq:Euler_equations} and
Grad~\eqref{eq:moment_equations_euler},~\eqref{eq:moment_equations_grad}
equations. This makes these equations naturally suitable for a variety
of boundary value problems, and, in particular, the Couette flow,
without having to resort to the the Fourier and Newton laws for the
stress and heat flux. In this work, we compute the Couette flow for
the conventional Navier-Stokes, diffusive Navier-Stokes, diffusive
Grad and regularized diffusive Grad equations, and compare the results
to the Direct Simulation Monte Carlo method \cite{Bird}.

\subsection{The DSMC method}

The Direct Simulation Monte Carlo (DSMC) method \cite{Bird} is a
``brute force'' approach to simulate a gas flow by computing the
motion and collisions of the actual gas molecules (thus the title of
the method). We must note, however, that the DSMC method does not
precisely simulate the exact molecular dynamics
in~\eqref{eq:dyn_sys_real}, and, in fact, the DSMC algorithm quite
closely resembles a coarse-grained version of the random jump process
in~\eqref{eq:dyn_sys_alt}. Indeed, in the DSMC algorithm the domain is
divided into the number of cells, and the collisions between the
molecules in a given cell are determined at random, based on their
number, size of the cell, the total collision cross-section, and the
molecular velocity. The outcomes of collisions are also generated at
random under the momentum and energy conservation constraints.

What makes the DSMC method somewhat more realistic than the random
jump process in~\eqref{eq:dyn_sys_alt}, is the presence of a collision
selection algorithm. In the random jump process, a collision occurs
whenever a random event arrives and the two molecules are within a
certain distance, whereas the DSMC method additionally takes into
account the velocities of molecules when computing the probabilities
of collisions. Thus, the DSMC method can be considered as a ``middle
of the road'' process between the realistic gas dynamics
in~\eqref{eq:dyn_sys_real} and the random jump process
in~\eqref{eq:dyn_sys_alt}. For this reason, the numerical simulations
below should be interpreted as a testing of the general sanity of the
proposed diffusive gas flow approximation; it is quite likely that the
empirical value of the scaled mass diffusivity $D_\alpha$, which we
use below, is specific to the DSMC solution, and may not necessarily
be suitable for a realistic gas flow.

Due to the detailed molecule interpretation of the gas, the DSMC
method allows great versatility of the molecule collision dynamics,
including the ability to simulate the collisions of polyatomic gas
molecules, and also the mixtures of different gases. The main
practical downside of the DSMC method is its tremendous computational
expense, as opposed to the fluid dynamics approach. For the DSMC
simulation, we test two different implementations of the DSMC method:
one is the DS1V\footnote{Available at
  \href{http://gab.com.au}{http://gab.com.au}} \cite{Bird}, and
another is the dsmcFoam\footnote{Part of the OpenFOAM software,
  \href{http://openfoam.org}{http://openfoam.org}}
\cite{ScaRooWhiDarRee}. We modified both the DS1V and dsmcFoam
software implementations to output the stress and heat flux inside the
domain, in addition to the density, velocity and temperature. Below we
demonstrate that the output of DS1V and dsmcFoam is nearly identical
for the cases we considered.

\subsection{Computation of the viscosity and mass diffusivity}

Away from the walls, for all fluid dynamics equations we used the
following expressions for the viscosity~$\mu$ and the scaled mass
diffusivity~$D_\alpha$:
\begin{equation}
\label{eq:mu_0_D_0}
\mu=\mu^*\sqrt{\frac{M\theta}{RT^*}},\qquad D_\alpha=D_\alpha^*\sqrt{
\frac{M\theta}{RT^*}},
\end{equation}
where $R=8.314$ kg m$^2$/(mol K sec$^2$) is the universal gas
constant, $M$ is the molar mass of the gas, $T^*$ is a constant
reference temperature specified in Kelvin units, and $\mu^*$
and~$D_\alpha^*$ are the reference values of the viscosity and scaled
mass diffusivity for $T^*$. Thus, rather than specifying the value of
the empirical scaling coefficient $\alpha$ from~\eqref{eq:Fext7}, we
instead specify the reference value $D_\alpha^*$ of the scaled
diffusion coefficient (which is, of course, also empirical).

Observe that both the viscosity $\mu$ and mass diffusivity $D$ (and,
therefore, its empirically scaled version $D_\alpha$) are proportional
to the mean free path of a gas molecule between collisions
\cite{ChaCow}. Previously in \cite{Abr15}, we computed the exact
scaling for the mean free path (and, therefore, viscosity) near a wall
under the assumption that the intermolecular collisions can be modeled
by a Poisson process, and subsequently found that the Knudsen boundary
layer appears in the Navier-Stokes velocity solution as a result of
the viscosity scaling. Here we use the same scaling for both the
viscosity $\mu$ and scaled mass diffusivity $D_\alpha$ near a wall:
\begin{equation}
\label{eq:mu_D_near_wall}
\frac{\mu^{\text{near wall}}}\mu=\frac{D_\alpha^{\text{near
      wall}}}{D_\alpha}=1+\frac 12\left(\frac x\lambda
E_1(x/\lambda)-e^{-x/\lambda}\right),
\end{equation}
where $x$ is the distance to the wall, $\lambda$ is the length of the
standard mean free path away from the wall, and $E_1(x)$ is the
exponential integral:
\begin{equation}
E_1(x)=\int_x^\infty\frac{e^{-y}}y\dif y.
\end{equation}
Observe that the scaling in~\eqref{eq:mu_D_near_wall} sets the second
derivatives of the velocity and temperature to infinity at the
wall~\cite{Abr15}, that is, the gas flow is formally always turbulent
at the wall. For the computation of $E_1(x)$ we use the approximation
proposed in \cite{SwaOjh}. To estimate the mean free path $\lambda$
from the thermodynamic quantities, we use the approximate formula
given in \cite{Cer2}, Chapter 5, eq. (1.3):
\begin{equation}
\label{eq:mfp}
\lambda=\frac\mu p\sqrt{\frac{\pi\theta}2}.
\end{equation}

\subsection{The Couette flow for argon}

\begin{figure}%
\includegraphics[width=\textwidth]{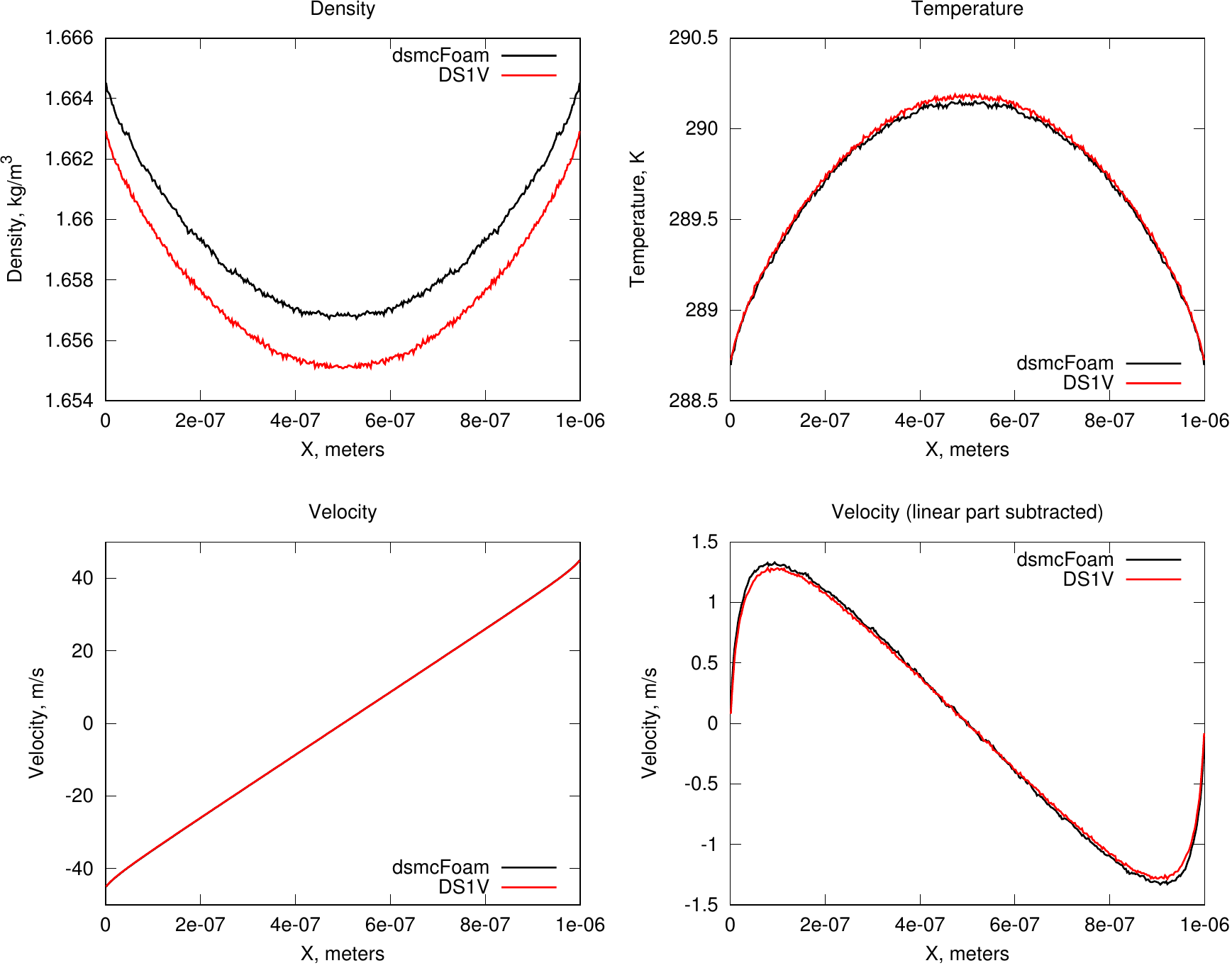}%
\caption{The density, velocity and temperature of the Couette flow for
  argon. The dsmcFoam compared to the DS1V.}%
\label{fig:argon_dsmcFoam_vs_DS1V}%
\end{figure}%
\begin{figure}%
\includegraphics[width=\textwidth]{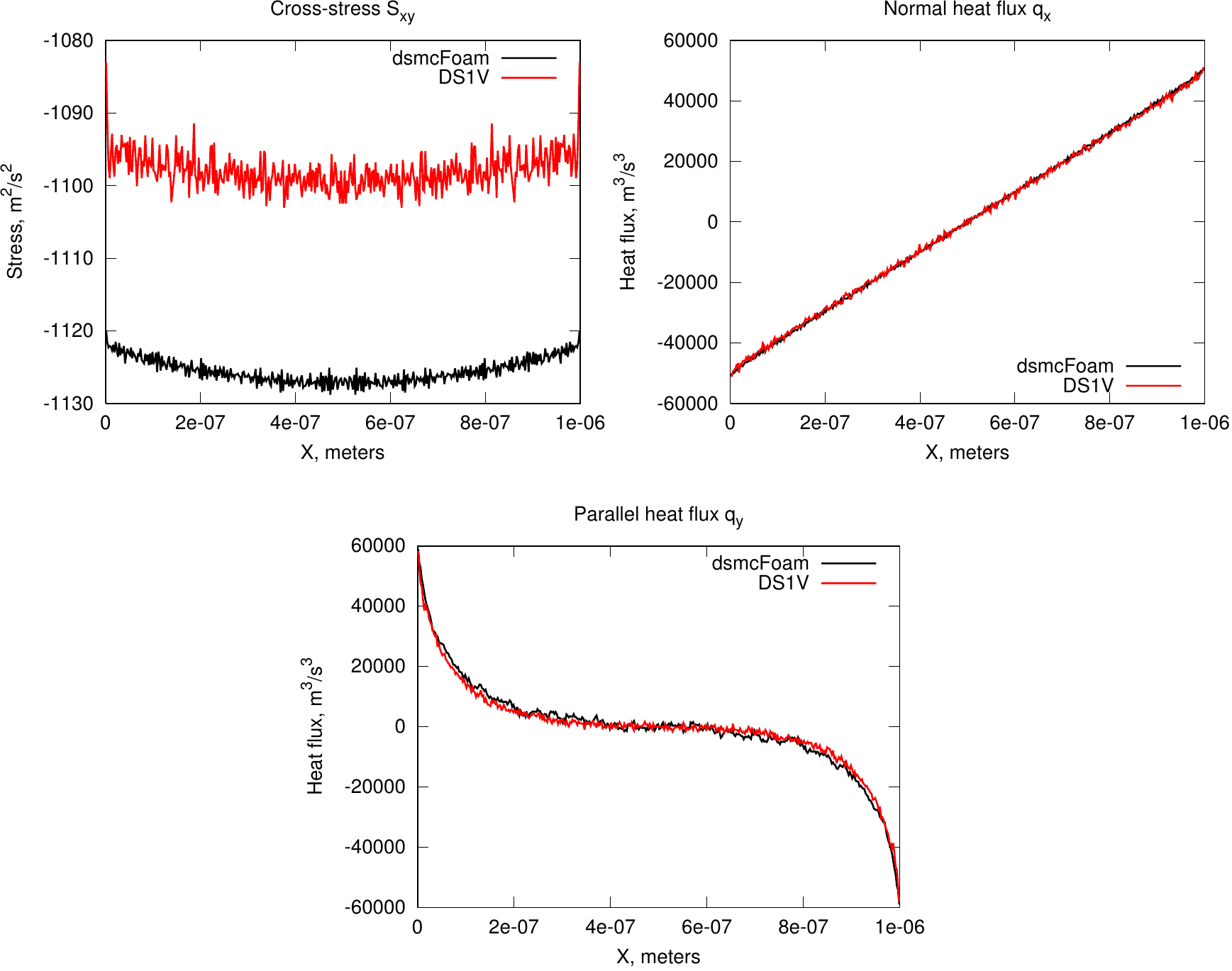}%
\caption{The cross-stress, normal heat flux, and parallel heat flux
  for the Couette flow for argon. The dsmcFoam compared to the DS1V.}%
\label{fig:argon_dsmcFoam_vs_DS1V_2}%
\end{figure}%
\begin{figure}%
\includegraphics[width=\textwidth]{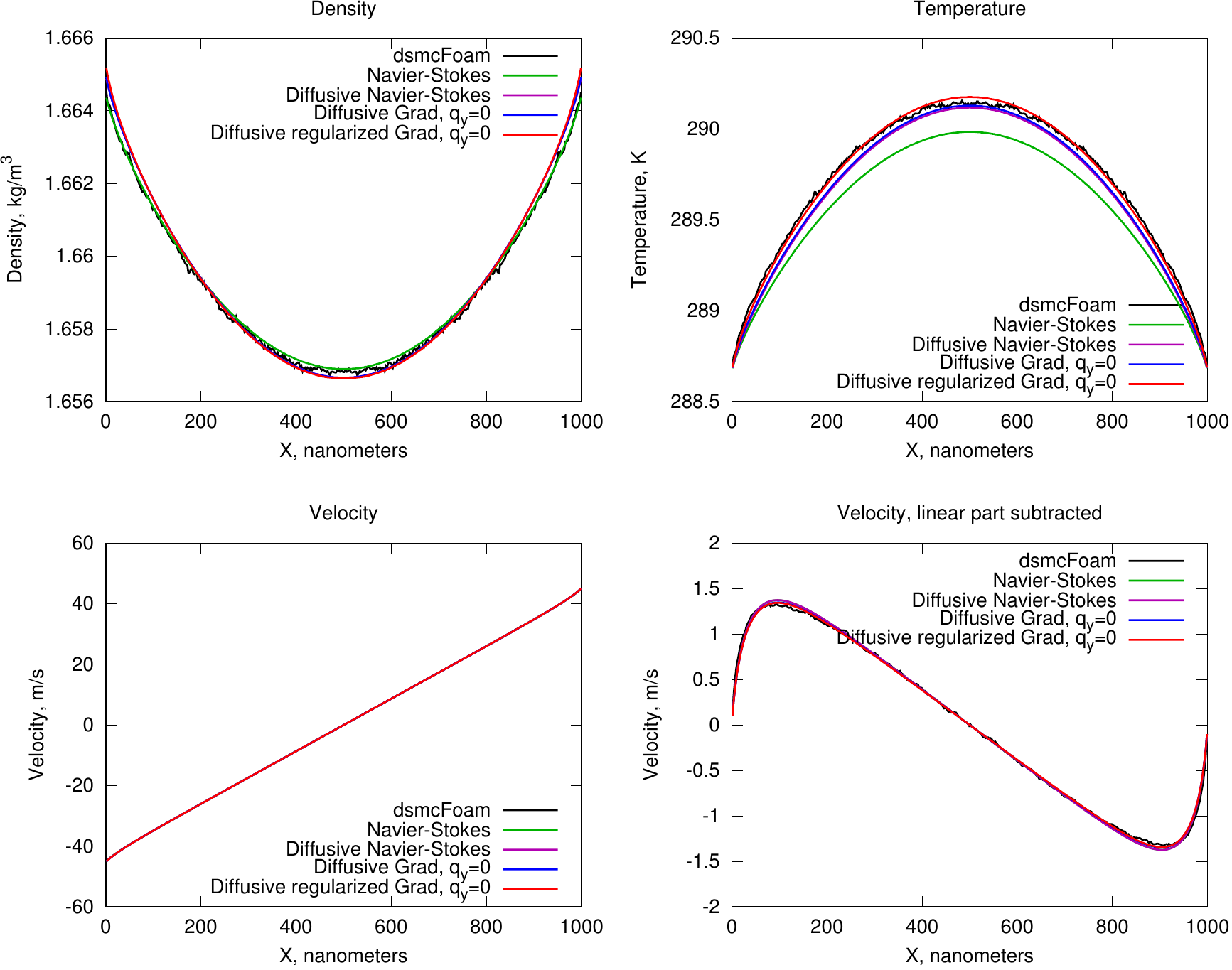}%
\caption{The density, velocity and temperature of the Couette flow for
  argon. The boundary parallel heat flux $q_y$ for the diffusive Grad
  and regularized diffusive Grad equations is set to zero.}%
\label{fig:argon}%
\end{figure}%
\begin{figure}%
\includegraphics[width=\textwidth]{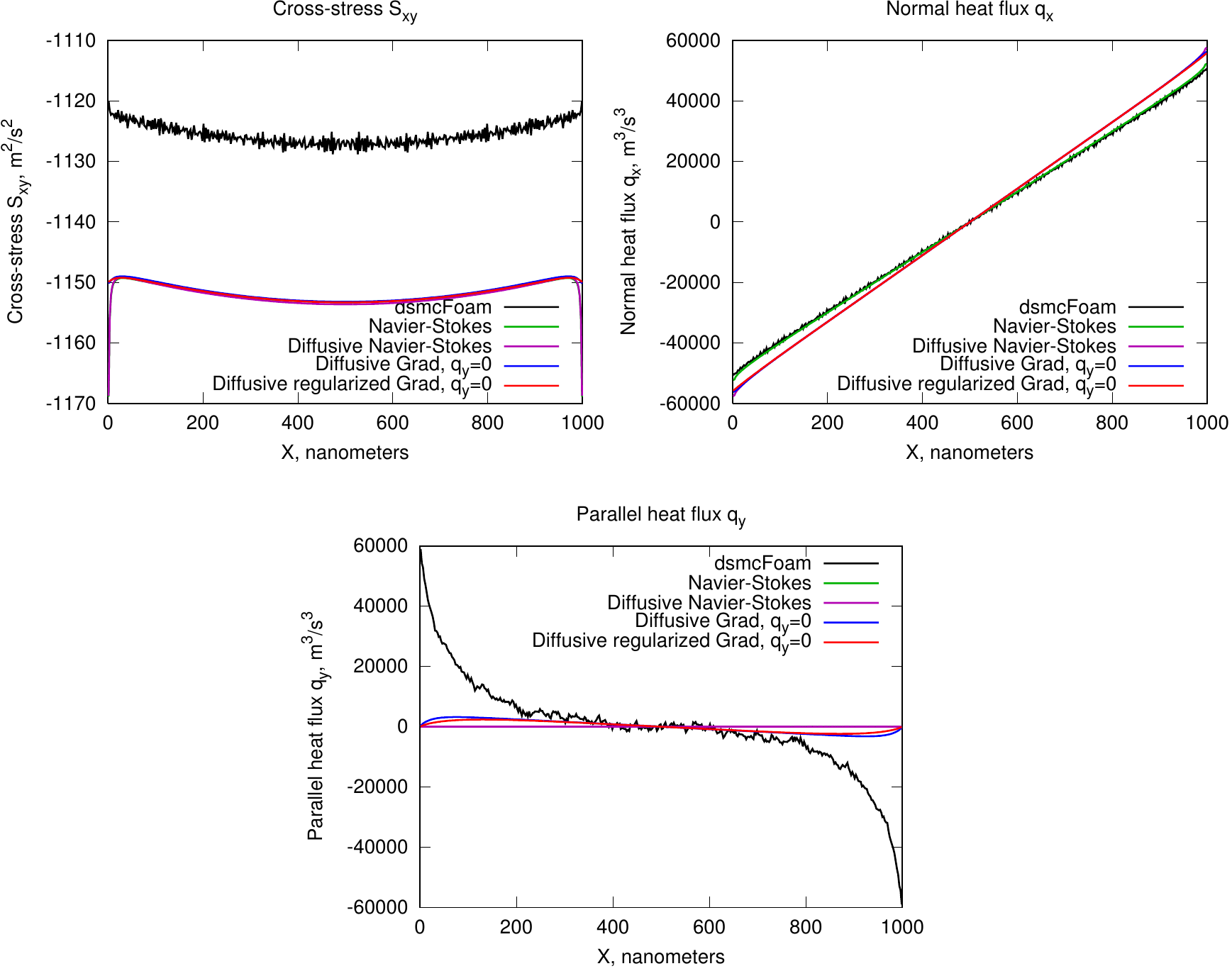}%
\caption{The cross-stress and heat flux for the Couette flow for
  argon. The boundary parallel heat flux $q_y$ for the diffusive Grad
  and regularized diffusive Grad equations is set to zero.}%
\label{fig:argon_2}%
\end{figure}%
\begin{figure}%
\includegraphics[width=\textwidth]{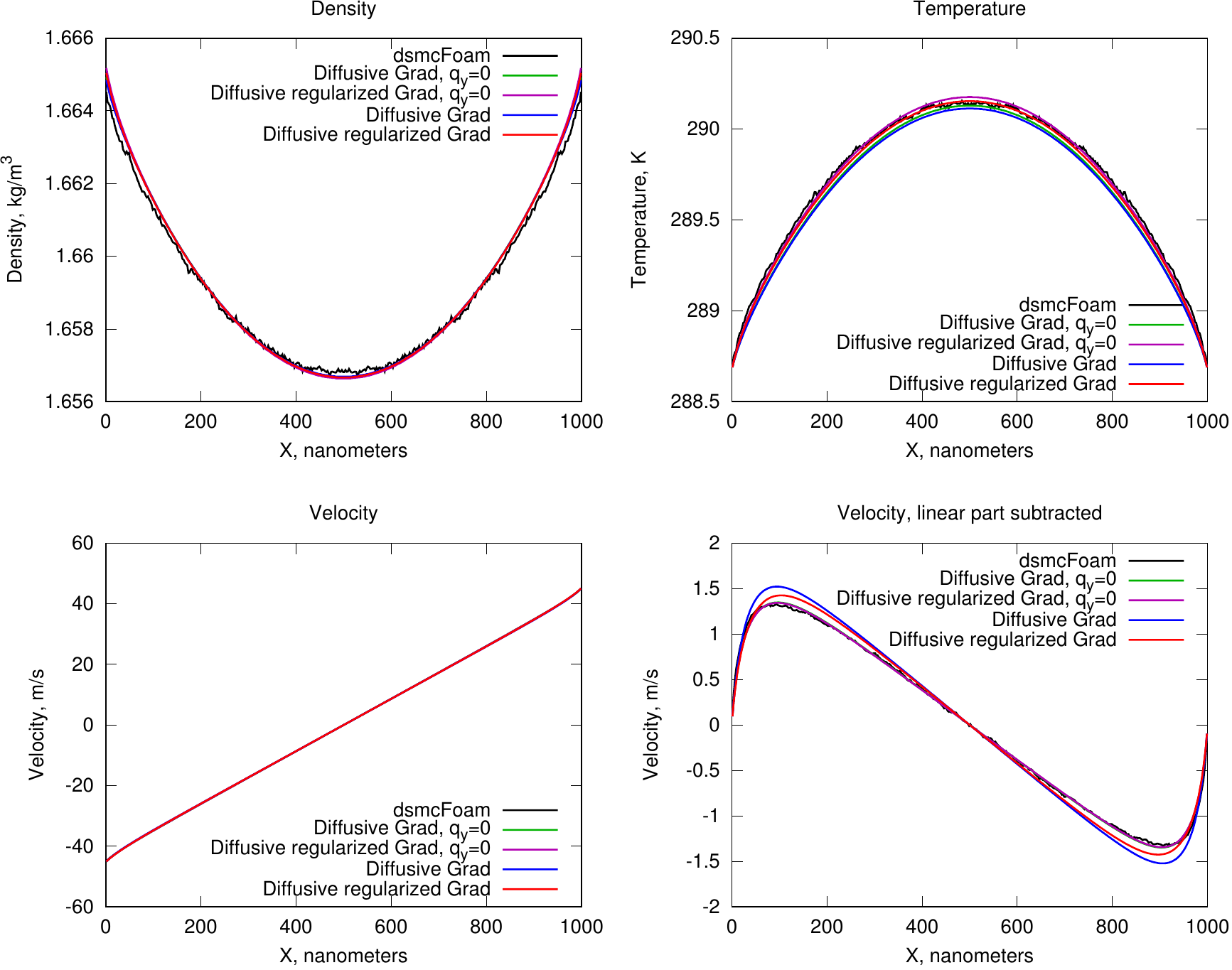}%
\caption{The density, velocity and temperature of the Couette flow for
  argon. Zero vs actual boundary heat flux $q_y$ for the diffusive Grad
and regularized diffusive Grad equations.}%
\label{fig:argon_3}%
\end{figure}%
\begin{figure}%
\includegraphics[width=\textwidth]{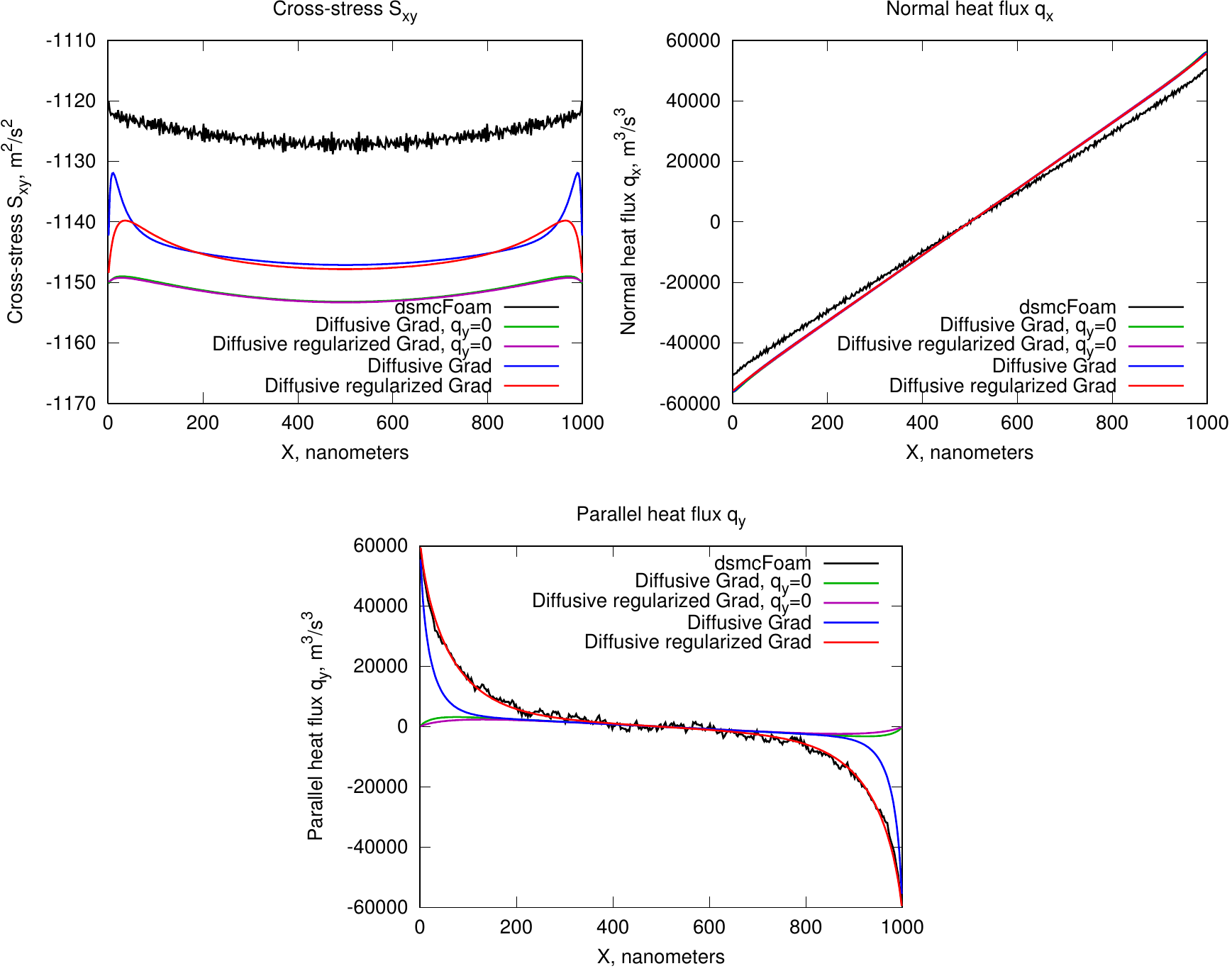}%
\caption{The cross-stress and heat flux for the Couette flow for
  argon. Zero vs actual boundary heat flux $q_y$ for the diffusive Grad
and regularized diffusive Grad equations.}%
\label{fig:argon_4}%
\end{figure}%

Argon is a monatomic gas, with observed kinetic behavior very close to
the ideal gas theory predictions ($\gamma=5/3$, $\Pran=2/3$). The DSMC
computational set-up for the Couette flow for argon was as follows:
\begin{itemize}
\item Distance between the walls: $10^{-6}$ meters;
\item Difference in wall velocities: 100 meters per second (the
  coordinate system is chosen so that the left wall moves at $-50$
  m/s, while the right wall moves at $50$ m/s);
\item Temperature of each wall: 288.15 K (15$^\circ$ Celsius);
\item Average number density of argon: $2.5\cdot 10^{25}$ molecules
  per cubic meter, which corresponds to the argon density $\rho\approx
  1.66$ kilogram per cubic meter. This number density is chosen so
  that the number of molecules is similar to that in the Earth
  atmosphere at sea level.
\end{itemize}
Due to slipping, the actual values of the thermodynamic quantities at
the boundaries were the following:
\begin{itemize}
\item Actual difference in the parallel velocity of the flow at the
  boundaries: 90.5 meters per second (which set the velocity of the
  gas flow at the boundaries at $\pm 45.25$ m/s);
\item Actual temperature of the flow at each boundary: 288.7 K.
\end{itemize}
The results of comparison of DS1V and dsmcFoam are shown in
Figure~\ref{fig:argon_dsmcFoam_vs_DS1V} for the density, velocity and
temperature, and in Figure~\ref{fig:argon_dsmcFoam_vs_DS1V_2} for the
stress and heat flux. Observe that the density, velocity and
temperature profiles in Figure~\ref{fig:argon_dsmcFoam_vs_DS1V} are
nearly identical (the slight difference in density is likely due to
the fact that slightly different numbers of molecules were simulated
by DS1V and dsmcFoam). The normal and parallel heat fluxes, shown in
Figure~\ref{fig:argon_dsmcFoam_vs_DS1V_2}, are also nearly identical.
The discrepancy in the cross-component of the stress in
Figure~\ref{fig:argon_dsmcFoam_vs_DS1V_2} is about 2.5\%. Below we use
the dsmcFoam as a benchmark for comparison against the fluid dynamics
simulations.

For all fluid dynamics equations we used the following parameters for
the viscosity and scaled mass diffusivity in~\eqref{eq:mu_0_D_0}: the
molar mass $M$ was set to $3.995\cdot 10^{-2}$ kg/mol for argon,
$T^*=288.15$ K (that is, 15$^\circ$ C), while the reference
constants~$\mu^*$ and $D_\alpha^*$ were chosen as follows:
\begin{itemize}
\item The reference viscosity~$\mu^*$ was set to $2.2\cdot 10^{-5}$
  kg/(m sec) at 15$^\circ C$ for argon \cite{HirCurBir,LemJac}.
\item The reference mass diffusion coefficient~$D_\alpha^*$ was chosen
  so that the diffusive Navier-Stokes equations for the Couette flow
  produced a good correspondence with the DSMC simulation (we found
  via a few trials that $D_\alpha^*=4\cdot 10^{-6}$ kg/(m sec)
  produces a good match).
\end{itemize}
We then carried out the numerical simulations with both the
conventional and diffusive Navier-Stokes equations until a steady
solution was reached, which we found to occur in about $1.5\cdot
10^{-7}$ seconds. The density, temperature and $y$-velocity profiles,
corresponding to both the conventional and diffusive Navier-Stokes
equations are compared with the corresponding DSMC profiles in
Figure~\ref{fig:argon}. Observe that the density and velocity profiles
are captured rather well by both the conventional and diffusive
Navier-Stokes equation (note that the velocity exhibits the Knudsen
boundary layer thanks to the scaling
in~\eqref{eq:mu_D_near_wall}). However, the temperature is
consistently underestimated by the conventional Navier-Stokes
equations, while the diffusive Navier-Stokes equations are more
accurate in this respect. We can see in Figure~\ref{fig:argon_2} that
the Newton and Fourier law approximations of the Navier-Stokes
equations for the cross-stress and normal heat flux develop
irregularities near the walls, which is likely due to the numerical
finite difference approximations of the derivatives in the presence of
the near-wall mean free path scaling. The parallel heat flux of both
the conventional and diffusive Navier-Stokes equations is zero, due to
the fact that the problem is translationally invariant along the
direction of the flow. Coincidentally, the underestimated temperature
of the conventional Navier-Stokes equations produces a better Fourier
law approximation to the normal heat flux of the DSMC solution, as
also shown in Figure~\ref{fig:argon_2}. Of course, one has to remember
that the DSMC method does not model the exact molecular dynamics
in~\eqref{eq:dyn_sys_real}, and thus the comparisons with an actual
measured gas flow need to be done for more definite conclusions.

For the diffusive Grad and regularized diffusive Grad equations,
observe that the stress and heat flux boundary conditions need to be
provided additionally. We found that setting these additional boundary
conditions directly to the values of the corresponding DSMC solution
leads to artificial boundary effects, possibly due to the fact that
the DSMC computation is not a solution of a partial differential
equation, and thus its boundary values are likely inconsistent with
the Grad equations. At the same time, the parallel heat flux from the
DSMC solution (which is identically zero for the conventional and
diffusive Navier-Stokes solutions) cannot simply be ignored.

To investigate this issue, we complete two sets of simulations. First,
we compute the solutions of the diffusive Grad and regularized
diffusive Grad equations with the boundary conditions for the stress
and heat flux chosen so that the corresponding Grad solution matches
the one of the diffusive Navier-Stokes equations, to verify that both
the diffusive Grad and regularized diffusive Grad equations indeed
approximate the diffusive Navier-Stokes equations for this regime. We
plot these solutions together with the Navier-Stokes solutions in
Figures~\ref{fig:argon} and~\ref{fig:argon_2}. Second, we compute the
solutions of the diffusive Grad and regularized diffusive Grad
equations with the parallel heat flux set to the DSMC value at the
boundary, and the rest of the boundary conditions left as described
above, to observe the effect of the nonzero parallel heat flux has on
the solution. These solutions are plotted in Figures~\ref{fig:argon_3}
and~\ref{fig:argon_4}, and compared against the zero parallel heat
flux Grad solutions.

\begin{figure}%
\includegraphics[width=\textwidth]{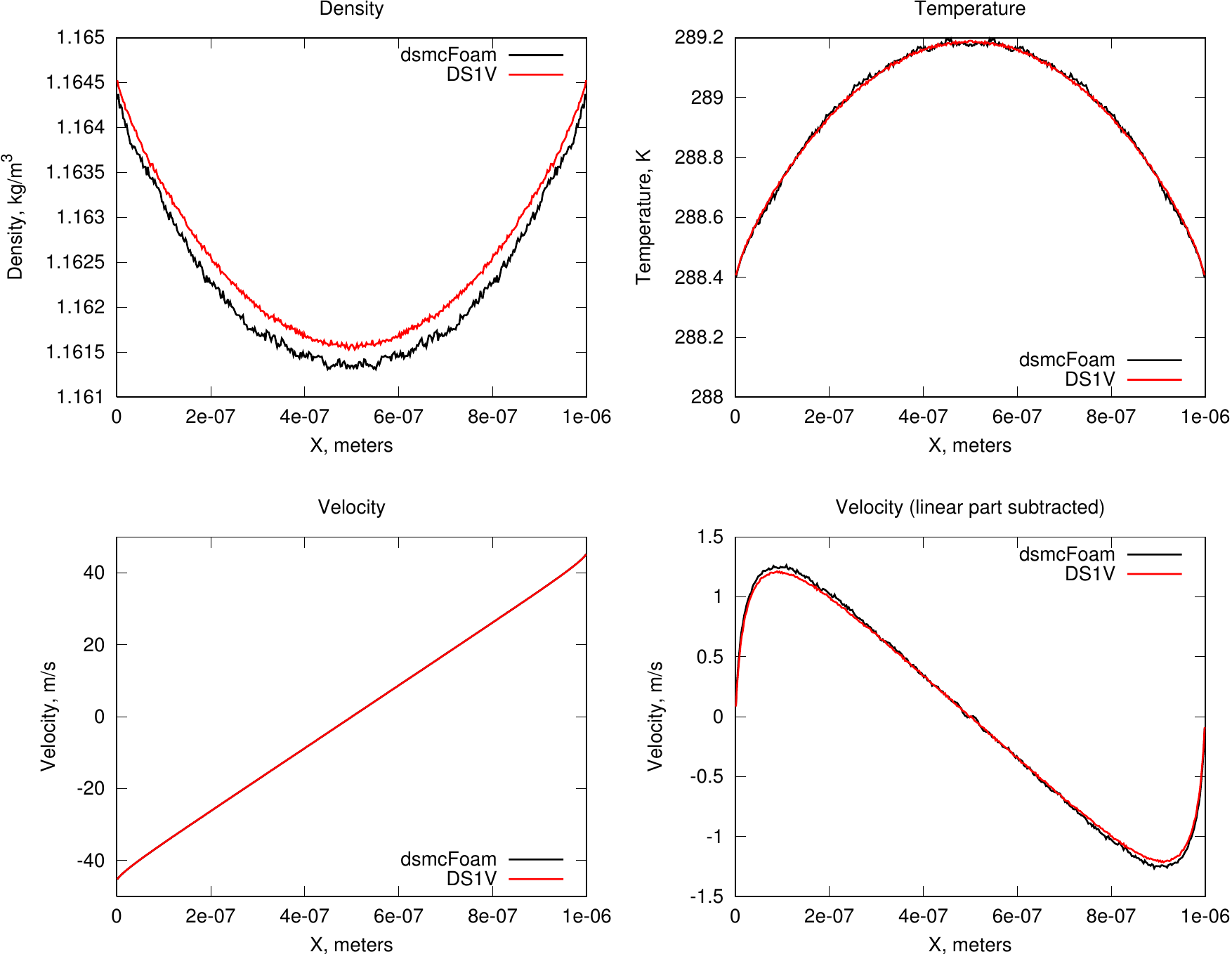}%
\caption{The density, velocity and temperature of the Couette flow for
  nitrogen. The dsmcFoam compared to the DS1V.}%
\label{fig:nitrogen_dsmcFoam_vs_DS1V}%
\end{figure}%
\begin{figure}%
\includegraphics[width=\textwidth]{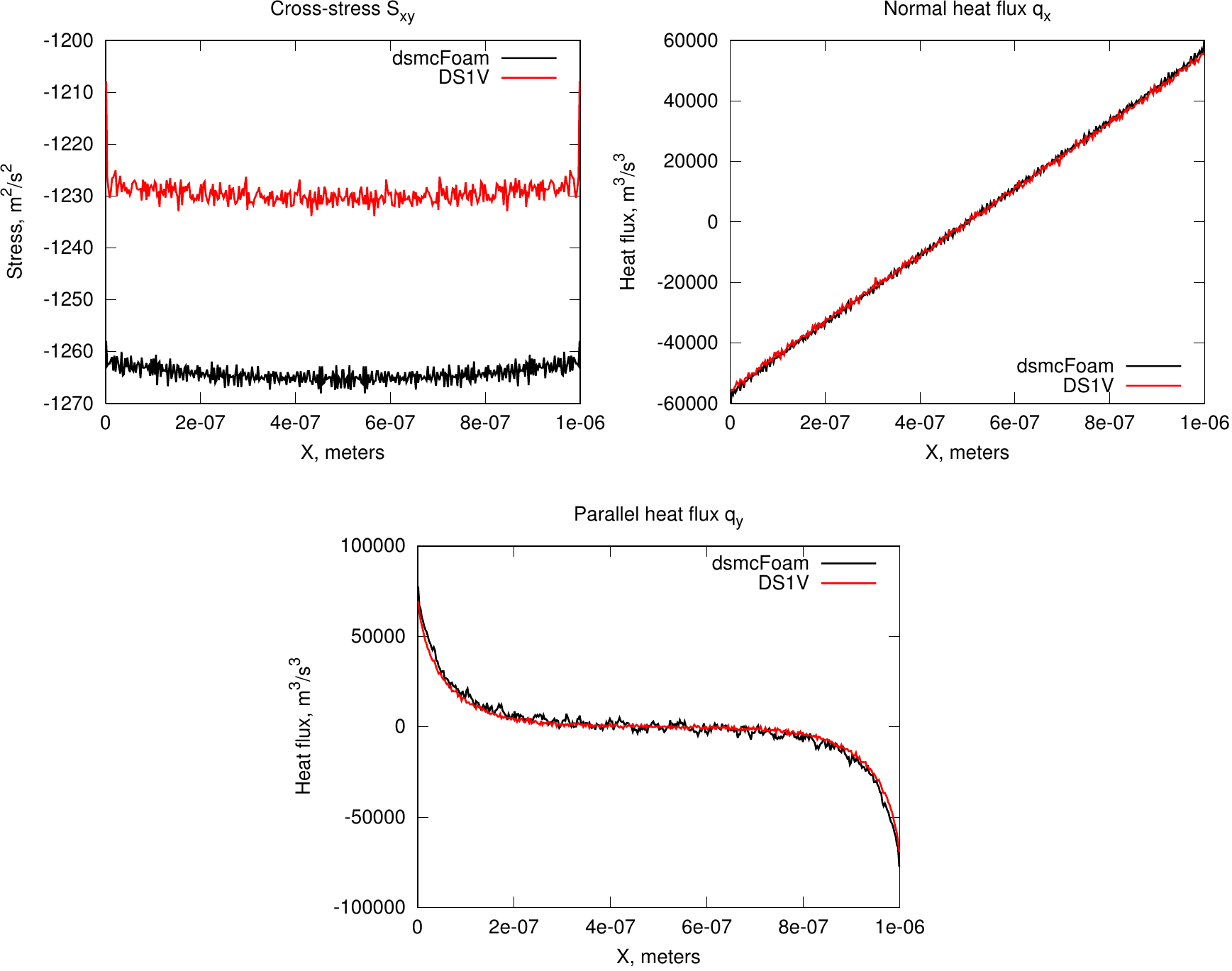}%
\caption{The cross-stress, normal heat flux, and parallel heat flux
  for the Couette flow for argon. The dsmcFoam compared to the DS1V.}%
\label{fig:nitrogen_dsmcFoam_vs_DS1V_2}%
\end{figure}%

Observe that the zero parallel heat flux Grad solutions in
Figures~\ref{fig:argon} and~\ref{fig:argon_2} are indeed very good
approximations to the diffusive Navier-Stokes equations away from the
walls, as all the computed variables are nearly identical. This also
suggests that the nonzero parallel heat flux in the DSMC computation
is strictly a boundary layer effect (unlike, for example, the normal
heat flux or cross-stress), which cannot be reproduced by the Newton
and Fourier approximations in~\eqref{eq:Sq_NS}. On the other hand,
setting the boundary value of the parallel heat flux to what was
computed by the DSMC results in somewhat different behavior between
the diffusive Grad and regularized diffusive Grad equations, which we
show in Figures~\ref{fig:argon_3} and~\ref{fig:argon_4}. While the
diffusive regularized Grad equations approximate the DSMC parallel
heat flux rather well, the diffusive Grad equations without the
regularization exhibit steeper fall-off of the parallel heat flux (see
Figure~\ref{fig:argon_4}). There is also a very slight ``improvement''
in the cross-stress component, however, note that it is smaller than
the difference between the DS1V and dsmcFoam computations, and thus
likely cannot be trusted with certainty. Also, while there is no
visibly discernible effect on the temperature or density, the velocity
Knudsen boundary layer is slightly ``de-tuned'', compared to the DSMC
solution, in both the diffusive Grad and regularized diffusive Grad
solutions (shown in Figure~\ref{fig:argon_3}). Given the tiny
magnitude of the change, it is, however, unclear whether the Knudsen
layer becomes less or more accurate -- first, the formula
in~\eqref{eq:mfp} does not necessarily output the exact value of the
free mean path, and, second, there is no guarantee that the Knudsen
layer of the actual measured gas flow is exactly the same as the one
in the DSMC computation.

\begin{figure}%
\includegraphics[width=\textwidth]{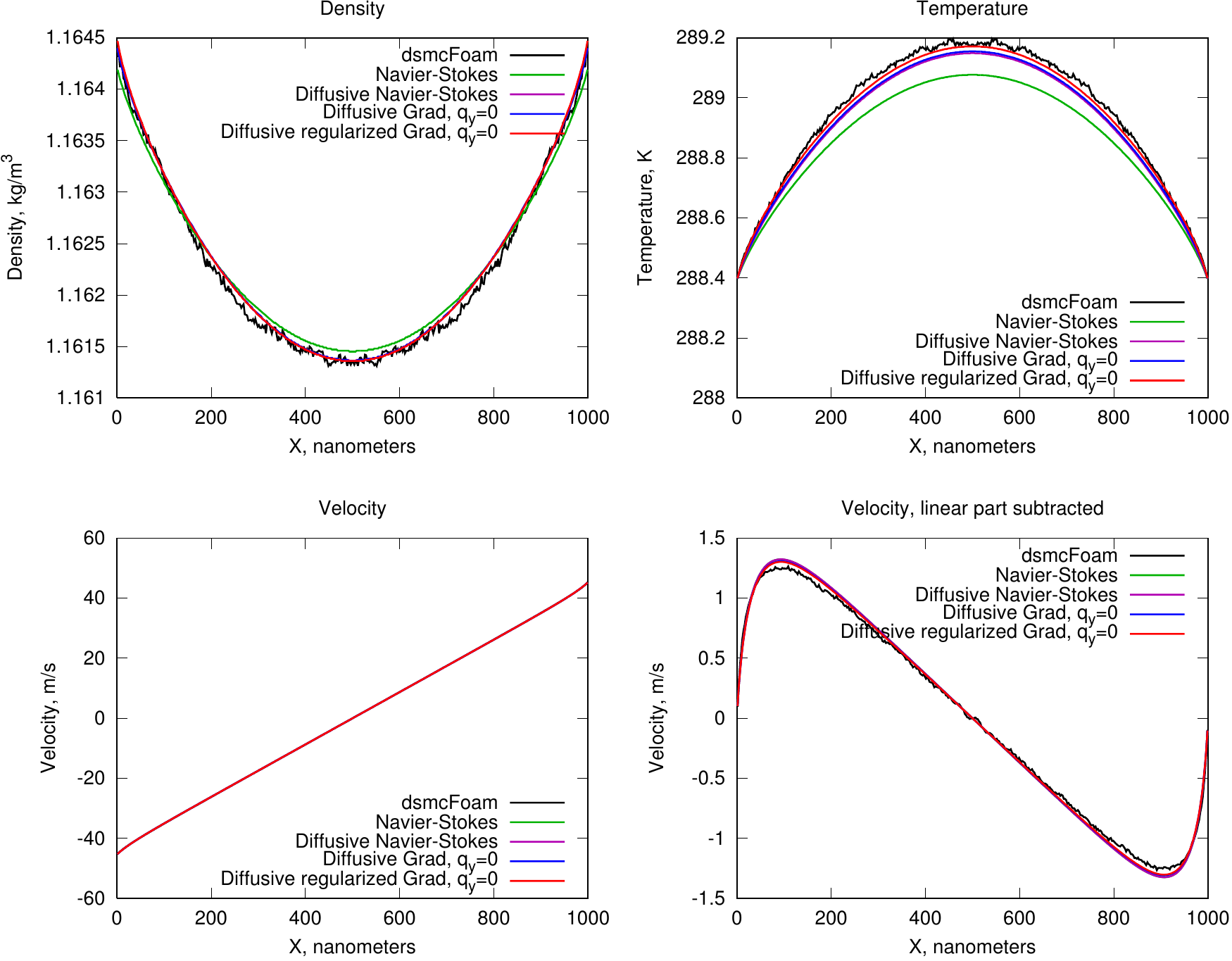}%
\caption{The density, velocity and temperature of the Couette flow for
  nitrogen. The boundary parallel heat flux $q_y$ for the diffusive
  Grad and regularized diffusive Grad equations is set to zero.}%
\label{fig:nitrogen}
\end{figure}%
\begin{figure}%
\includegraphics[width=\textwidth]{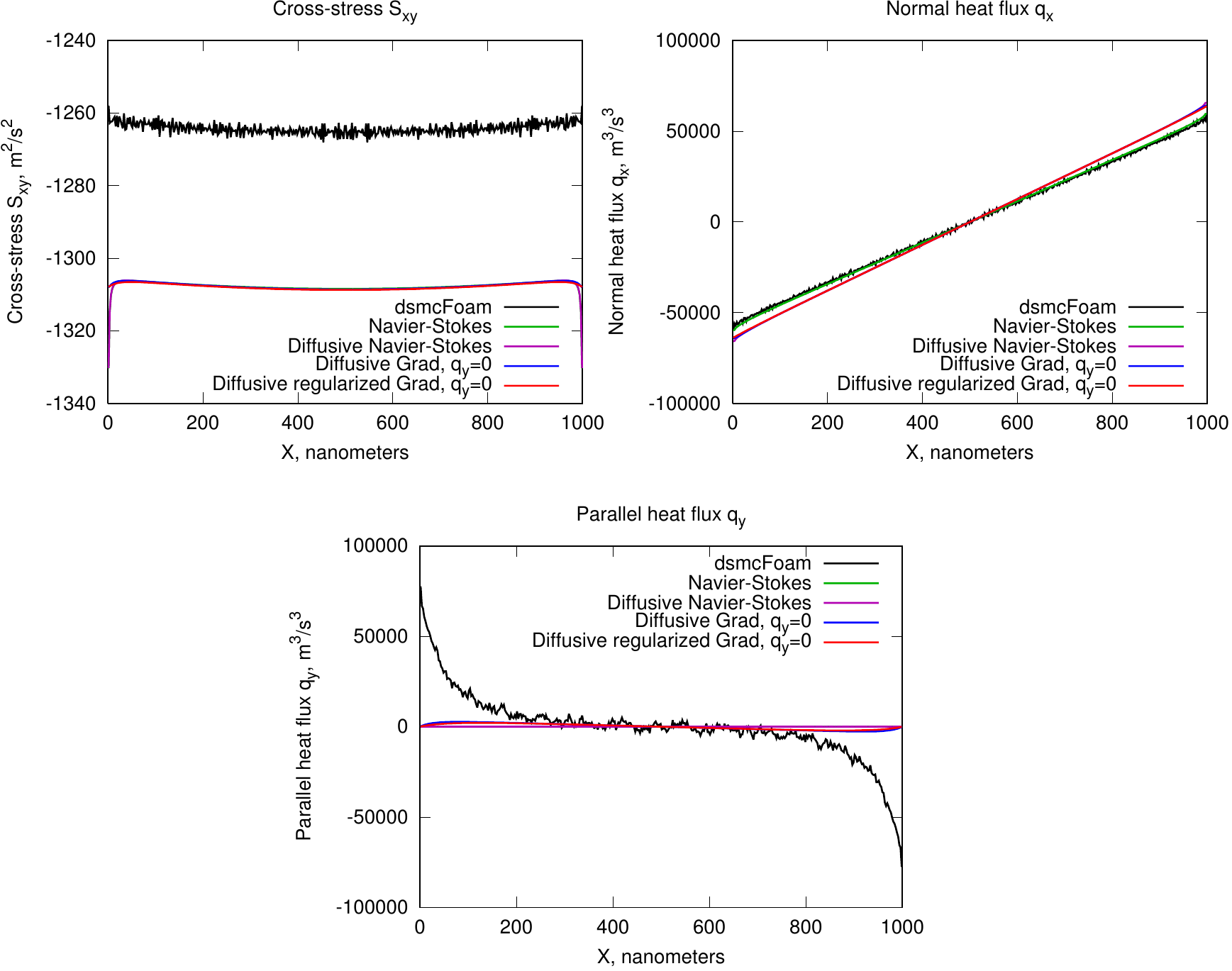}%
\caption{The cross-stress and heat flux for the Couette flow for
  nitrogen. The boundary parallel heat flux $q_y$ for the diffusive
  Grad and regularized diffusive Grad equations is set to zero.}%
\label{fig:nitrogen_2}
\end{figure}%
\begin{figure}%
\includegraphics[width=\textwidth]{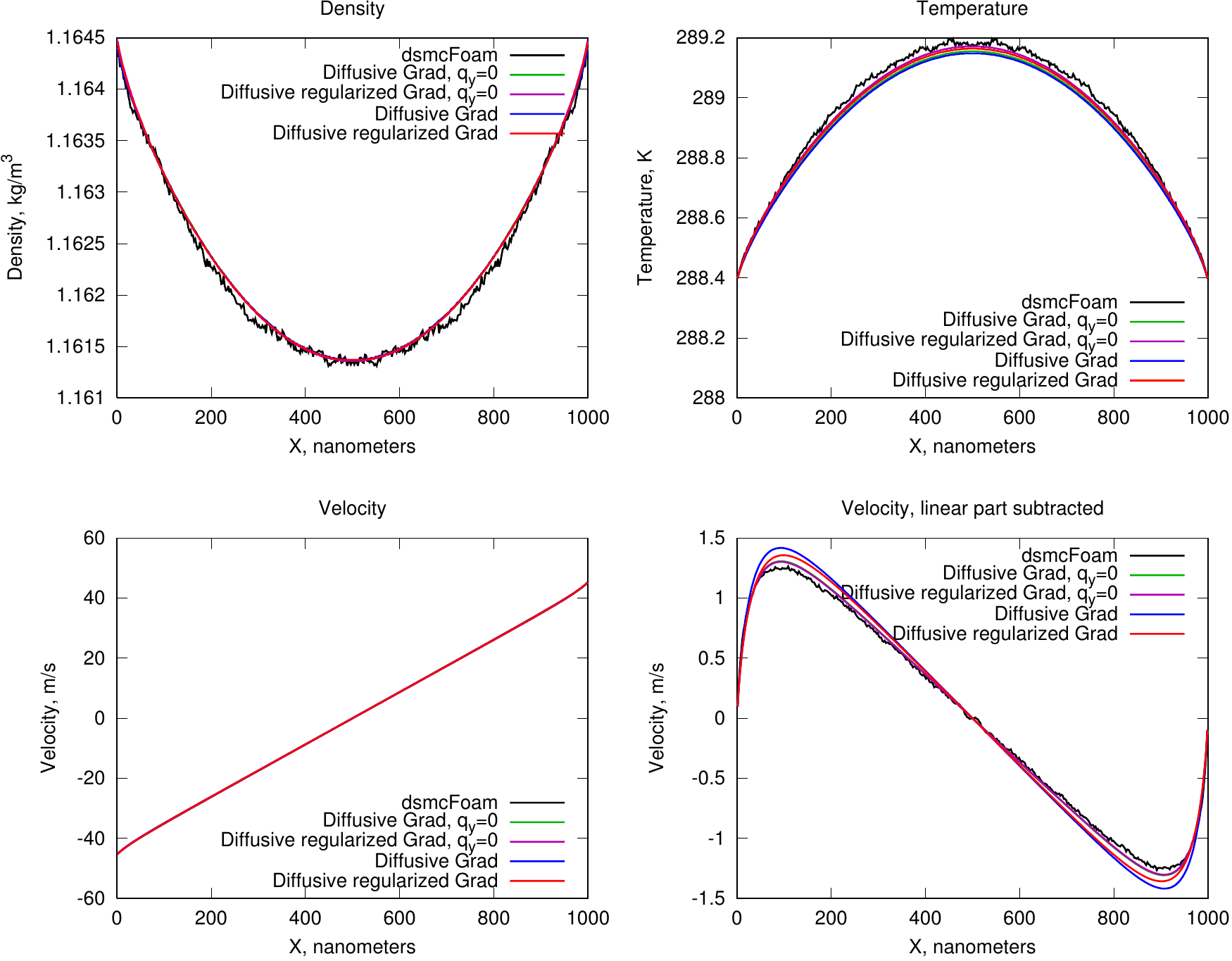}%
\caption{The density, velocity and temperature of the Couette flow for
nitrogen. Zero vs actual boundary heat flux $q_y$ for the diffusive Grad
and regularized diffusive Grad equations.}%
\label{fig:nitrogen_3}
\end{figure}%
\begin{figure}%
\includegraphics[width=\textwidth]{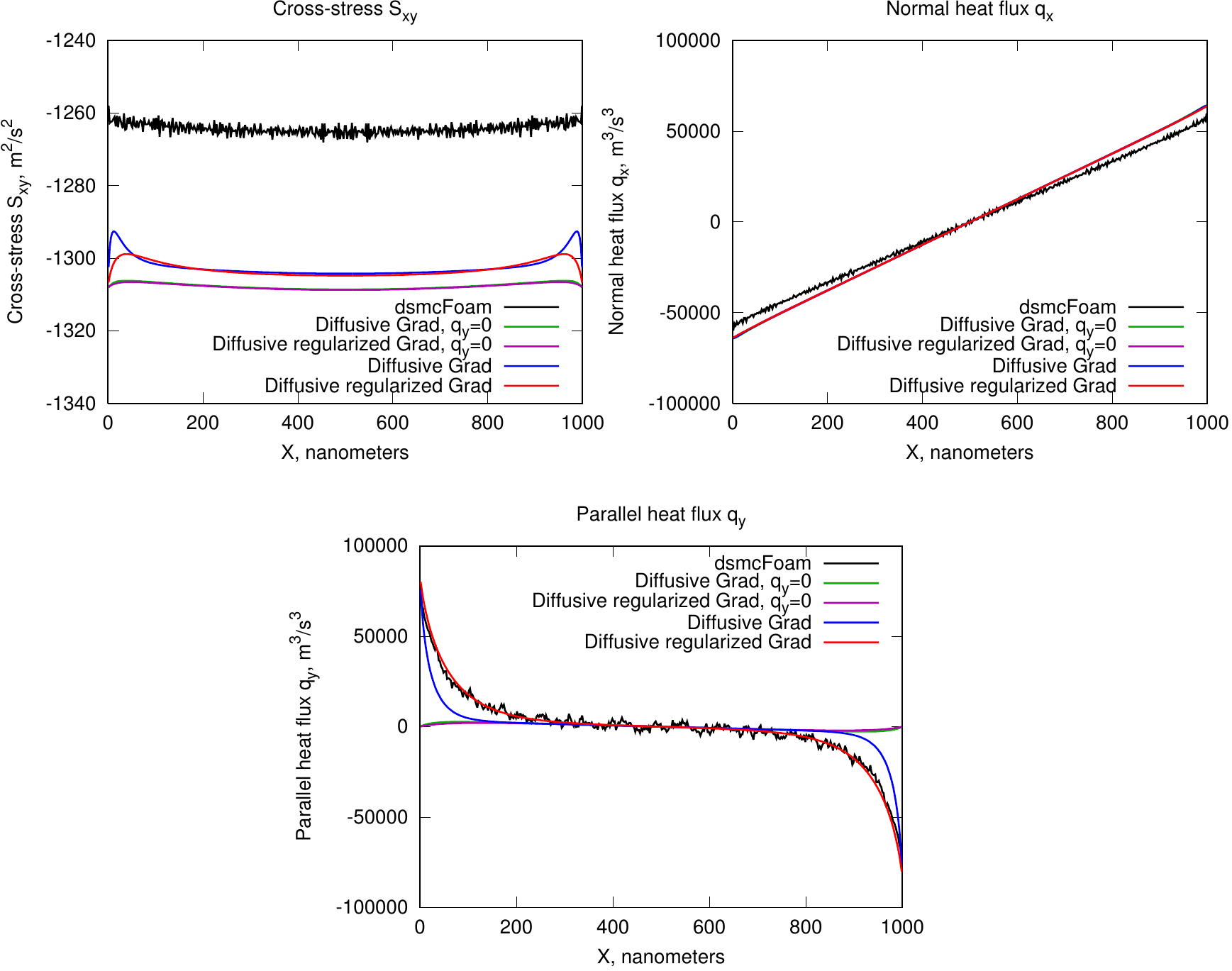}%
\caption{The cross-stress and heat flux for the Couette flow for nitrogen.
Zero vs actual boundary heat flux $q_y$ for the diffusive Grad
and regularized diffusive Grad equations.}%
\label{fig:nitrogen_4}
\end{figure}%

Generally, we conclude from Figures \ref{fig:argon}--\ref{fig:argon_4}
that if one does not require the accuracy of the parallel heat flux
near the walls, then the diffusive Navier-Stokes equations are already
quite accurate for all the other variables, and there is no need to
upgrade the model to the nonequilibrium Grad closure. If, however, one
requires an accurate approximation for the parallel heat flux, then the
regularized diffusive Grad equations should be used, as the diffusive
Grad equations without regularization produce the parallel heat flux
with a significantly steeper fall-off away from the wall. It appears
that the main practical strength of the diffusive and regularized
diffusive Grad equations could manifest in situations where
significant external heat fluxes are present in the system (for
example, modeling the Earth atmosphere with the external surface heat
flux appears to be such an application).

\subsection{The Couette flow for nitrogen}

The DSMC computational set-up for the Couette flow for nitrogen was
largely the same as previously for argon:
\begin{itemize}
\item Distance between the walls: $10^{-6}$ meters;
\item Difference in wall velocities: 100 meters per second;
\item Temperature of each wall: 288.15 K (15$^\circ$ Celsius);
\item Average number density of nitrogen: $2.5\cdot 10^{25}$ molecules
  per cubic meter (which corresponds to the nitrogen density $\rho=1.16$
  kilogram per cubic meter).
\end{itemize}
Due to slipping, the actual values of the thermodynamic quantities at
the boundaries were the following:
\begin{itemize}
\item Actual difference in the parallel velocity of the flow at the
  boundaries: 90.9 meters per second;
\item Actual temperature of the flow at each boundary: 288.4 K.
\end{itemize}
The results of comparison of DS1V and dsmcFoam are shown in
Figure~\ref{fig:nitrogen_dsmcFoam_vs_DS1V} for the density, velocity
and temperature, and in Figure~\ref{fig:nitrogen_dsmcFoam_vs_DS1V_2}
for the stress and heat flux. Just like previously for argon, here
observe that the density, velocity and temperature profiles in
Figure~\ref{fig:argon_dsmcFoam_vs_DS1V} are nearly identical (the
slight difference in density is likely due to the fact that slightly
different numbers of molecules were simulated by DS1V and
dsmcFoam). The normal and parallel heat fluxes, shown in
Figure~\ref{fig:argon_dsmcFoam_vs_DS1V_2}, are also nearly identical.
The discrepancy in the cross-component of the stress in
Figure~\ref{fig:argon_dsmcFoam_vs_DS1V_2} is about 2.5\%. As
previously for argon, here we use the dsmcFoam as a benchmark for
comparison against the fluid dynamics computations.

For both the conventional and diffusive the Navier-Stokes equations we
used the same expressions for the viscosity~$\mu$ and the scaled mass
diffusivity coefficient~$D_\alpha$ as in~\eqref{eq:mu_0_D_0} with the
same reference diffusivity constant $D_\alpha^*=4\cdot 10^{-6}$ kg/(m
sec), however, the molar mass~$M$ was replaced with that of nitrogen
(that is, $2.801\cdot 10^{-2}$ kg/mol) and the reference viscosity
constant was set to $1.74\cdot 10^{-5}$ kg/(m sec), which is a
standard value for nitrogen at $15^\circ$~C \cite{HirCurBir,LemJac}.
The Prandtl number was set to $\Pran=0.69$ (also a standard value for
nitrogen at 15$^\circ$ C).  We then carried out numerical simulations
with both the conventional and diffusive Navier-Stokes equations until
a steady solution was reached (also about $1.5\cdot 10^{-7}$
seconds). The density, temperature and $y$-velocity profiles,
corresponding to both the conventional and diffusive Navier-Stokes
equations, are compared with the DSMC profiles on
Figure~\ref{fig:nitrogen}. Observe that, just as above for argon, the
density and velocity profiles are captured rather well by both the
conventional and diffusive Navier-Stokes equations, including the
Knudsen boundary layers for the velocity. However, the temperature is
consistently underestimated by the conventional Navier-Stokes
equations, while the diffusive Navier-Stokes equations are somewhat
more accurate. Again, coincidentally, the underestimated temperature
of the conventional Navier-Stokes equations produces a better
approximation to the normal heat flux of the DSMC solution, as shown
in Figure~\ref{fig:nitrogen_2}. The parallel heat flux of both the
conventional and diffusive Navier-Stokes equations is zero, while both
the cross-stress and normal heat flux develop irregularities near the
walls, likely due to finite difference approximations as above for
argon.

For the diffusive Grad and regularized diffusive Grad equations, we
complete the two sets of simulations just as above for argon, first
one with the boundary values for the stress and heat flux to match the
solution of the diffusive Navier-Stokes equations (shown in
Figures~\ref{fig:nitrogen} and~\ref{fig:nitrogen_2}), and the second
one with the parallel heat flux set to the DSMC value at the boundary
(shown in Figures~\ref{fig:nitrogen_3} and~\ref{fig:nitrogen_4}).
Just as for argon, the zero parallel heat flux Grad solutions shown in
Figures~\ref{fig:nitrogen} and~\ref{fig:nitrogen_2} accurately match
the solution of the diffusive Navier-Stokes equations away from the
walls. However, when the boundary value of the parallel heat flux is
set to what was computed by the DSMC, the diffusive regularized Grad
equations approximate the DSMC parallel heat flux rather well, while
the diffusive Grad equations without the regularization exhibit
steeper fall-off of the parallel heat flux (see
Figure~\ref{fig:nitrogen_4}), just as for argon above. The velocity
Knudsen boundary layer is again slightly changed, as compared to the
DSMC solution, in both the diffusive Grad and regularized diffusive
Grad solutions (shown in Figure~\ref{fig:nitrogen_3}).

Generally, the results in Figures
\ref{fig:nitrogen}--\ref{fig:nitrogen_4} match those for argon in
Figures \ref{fig:argon}--\ref{fig:argon_4}, despite the fact that
nitrogen is a gas with a qualitatively different behavior than argon
(diatomic, the trace of the stress matrix is nonzero and thus requires
a fourteenth equation in the Grad closure). Given the fact that we
also re-used the argon values of the higher-order Prandtl numbers for
the third and fourth moments in the Grad regularization expressions
in~\eqref{eq:Reg_Grad} and~\eqref{eq:Reg_Grad_2} for nitrogen, the
accuracy of the results for nitrogen is quite surprising. It remains
to be seen whether the regularized diffusive Grad equations could
potentially be used in the modeling of Earth atmosphere in the
presence of strong external heat fluxes.

\section{Summary}
\label{sec:summary}

In this work we develop a spatially diffusive analog of the Boltzmann
equation, based on the difference between the realistic gas dynamics
and the random motion which is modeled by the Boltzmann equation. For
that, we first construct a precise multimolecular random process
which, in the appropriate limit, leads to the Boltzmann
equation. Next, we apply the standard multiscale expansion formalism
to the difference dynamics between the realistic gas and the
constructed random jump process, and compute the long-term
homogenization dynamics for the difference coordinate of a molecule in
the form of a diffusion process. We adjust the constructed
multimolecular random jump process with this diffusion process, which
leads to the Boltzmann equation with an additional spatial diffusion
term. We then obtain the hierarchy of the diffusive moment equations
from the Boltzmann equation in a standard way, and carry out a
computational study of the both the conventional and diffusive
Navier-Stokes equations, as well as diffusive and regularized
diffusive Grad closures of the moment equations in a simple Couette
flow setting with argon and nitrogen. We compare the results with the
Direct Simulation Monte Carlo computations. We find that all studied
moment closures develop the full-fledged Knudsen velocity boundary
layers near the walls, closely matching the results of the DSMC
computations. We also note that the conventional Navier-Stokes
equations tend to underestimate the temperature away from the walls,
while the diffusive Navier-Stokes and Grad closures are more accurate
in this respect. Additionally, we find that the component of the heat
flux parallel to the flow, produced by the DSMC computations, is
captured quite well by the diffusive regularized Grad equations.

For the future study, the natural step forward is to investigate the
behavior of the new equations in capillary gas flows under normal
conditions, as well as rarefied gas flows, in more advanced spatial
configurations. One of the advantages of the new equations is that
they combine the ability to model the flows of polyatomic gases (which
are ubiquitous in nature) with the higher-order Grad closure, since,
to our knowledge, thus far the Grad closure dynamics, where used, were
confined to a monatomic set-up. From the kinetic theory perspective,
an interesting problem is to estimate the value of the empirical
coefficient $\alpha$ in the diffusive scaling~\eqref{eq:D} from the
basic principles.

Also, we plan to investigate whether the new equations can be used for
modeling turbulent flows in the presence of strong external heat
fluxes, such the large scale atmospheric circulation. It is likely
that the major benefits of the regularized diffusive Grad
approximation should manifest in situations where the heat fluxes are
important, since we observed above that the Grad closure demonstrates
the ability of capturing both normal and parallel heat fluxes (unlike
the Navier-Stokes closures, which can only capture the normal heat
flux in the studied set-up). An attractive feature of the diffusive
Grad closure is that it allows to prescribe the stress and heat flux
at the boundaries explicitly, which could lead to a more detailed
model of the energy exchange between the atmosphere and the surface of
Earth.

\ack The author thanks Ibrahim Fatkullin for interesting discussions,
and Jasmine Otto for the help with the OpenFOAM and dsmcFoam software.
The work was supported by the Office of Naval Research grant
N00014-15-1-2036.

\end{document}